\documentclass[12pt]{iopart}
\usepackage{graphicx}% Include figure filestau
\usepackage{dcolumn}% Align table columns on decimal point
\usepackage{bm}% bold math
\usepackage[dvips]{color}
\usepackage{epsfig}

\def\la{\mathrel{\mathpalette\fun <}}
\def\ga{\mathrel{\mathpalette\fun >}}
\def\fun#1#2{\lower3.6pt\vbox{\baselineskip0pt\lineskip.9pt
  \ialign{$\mathsurround=0pt#1\hfil##\hfil$\crcr#2\crcr\sim\crcr}}}

\def\erf{\mathop{\rm erf}}

\newcommand{\ie}{{\it i.e.}}

\def\nab{{\hbox{\boldmath$\nabla$}}}

\definecolor{Black}{named}{Black}
\definecolor{Blue}{named}{Blue}
\definecolor{Red}{named}{Red}
\definecolor{Green}{named}{Green}
\definecolor{Brown}{named}{Brown}

\begin{document}

\begin{flushright}
{\large \tt DESY 10-131}
\end{flushright}

\title[]{Inverse Compton gamma-rays from Galactic dark matter annihilation:
 Anisotropy signatures}
\author{Le Zhang$^1$, Francesco Miniati$^2$ and G\"unter Sigl$^1$} 
\address{$^1$ II. Institut f\"ur theoretische Physik, Universit\"at Hamburg,
Luruper Chaussee 149, D-22761 Hamburg, Germany} 
\address{$^2$ Physics Department, Wolfgang-Pauli-Strasse 27, ETH Z\"urich, CH-8093 Z\"urich, Switzerland}

\begin{abstract}
  High energy electrons and positrons from annihilating dark matter
  can imprint unique angular anisotropies on the diffuse gamma-ray
  flux by inverse Compton scattering off the interstellar radiation field. We develop a numerical tool to compute gamma-ray emission
  from such electrons and positrons produced
 in the smooth host halo
  and in substructure halos with masses down to $10^{-6}M_\odot$.  
We show that 
%, unlike the total gamma-ray angular power spectrum observed by Fermi-LAT,
 the angular power spectrum from inverse
  Compton scattering is exponentially suppressed below an angular
  scale determined by the diffusion length of electrons and
  positrons.
% For TeV scale dark matter with a canonical thermal
%  freeze-out cross section $3\times10^{-26} {\rm cm^3/s}$, this
%  feature may be detectable by Fermi-LAT in the energy range 100-300
%  GeV after 
%more 
%sophisticated foreground subtraction. 
 We also find
  that the total flux and the shape of the angular power spectrum
  depends sensitively on the spatial distribution of subhalos in the
  Milky Way. Finally, the contribution from the smooth host halo
  component to the gamma-ray mean intensity is negligibly small
  compared to subhalos.
\end{abstract}

\maketitle

\section{Introduction} \label{sec:intro} The existence of nonbaryonic
dark matter in modern cosmology is strongly supported by several
independent signatures~\cite{Bertone:2004pz}, including the cosmic
microwave background (CMB), gravitational lensing and large scale
structure surveys. However, the nature of dark matter remains a
mystery. To reproduce the correct relic density, it is naturally
assumed that dark matter is composed of weakly interacting massive
particles (WIMPs) such as the supersymmetric neutralino which is one
of the most popular candidates~\cite{Jungman1996,Hooper2007a}. Since
the self-annihilation rate is proportional to the dark matter number
density squared, potential signals from dark matter annihilation are
most likely to be detected in highly dense regions such as the centers
of galaxies. It is also expected that the energy released in dark
matter annihilation in the early Universe can influence CMB
anisotropies. With the high precision CMB data from the WMAP
satellite~\cite{Komatsu05}, strong constraints on the properties of
dark matter can be
obtained~\cite{Chen:2003gz,Zhang:2007zzh,Zhang:2006fr}. A
multi-wavelength analysis of dark matter annihilations has been
proposed recently in Ref.~\cite{Colafrancesco:2005ji,Profumo:2010ya}.

Recently, several cosmic ray and gamma-ray experiments provided new
windows to detect the signatures of dark matter. In
particular, the PAMELA satellite~\cite{Adriani:2008zr} reported an
``excess'' of the positron fraction above 10 GeV. Also the $e^-+e^+$
spectra above a few hundred GeV from the Fermi-LAT~\cite{Abdo:2009zk}
and HESS measurements~\cite{Collaboration:2008aaa} are significantly
harder than expected~\cite{Grasso:2009ma,DiBernardo:2009iu}. These
striking observations were often interpreted in terms of dark
matter annihilation or decay which can contribute to the flux of high
energy electrons and
positrons~\cite{Cirelli:2008pk,ann1,ann2,ann3,ann4,Bai:2009ka}.
 However, standard astrophysical sources, such as
pulsars~\cite{Malyshev:2009tw,Profumo:2008ms} and supernova
remnants~\cite{Kobayashi:2003kp,Shaviv:2009bu,Blasi:2009hv} also
provide a possible explanation for these excesses.

The high energy electrons and positrons induced by WIMP annihilation
%during their propagation can 
produce gamma-rays through 
%the process of
inverse Compton scattering (ICS) off the low energy background photons
in the interstellar radiation field (ISRF) and through bremsstrahlung
emissions due to the interaction with the ionized interstellar medium.
These components add to the prompt gamma-ray radiation accompanying
pairs emitted in the annihilation
event~\cite{Bergstrom:2004cy}. Gamma-ray dark matter
  signatures could be identified by the Fermi Large Area
  Telescope~\cite{Atwood2009}, due to its unprecedented angular and
  energy resolution, despite the fact that the gamma-ray spectrum is
  dominated by conventional astrophysical sources, such as pulsars,
  supernovas, blazars~\cite{chiang98,stecker96} and perhaps structure
  formation shocks~\cite{miniati02}. Many
  efforts~\cite{Baltz2008,Cirelli:2009vg,Baltz:2004bb,Yuan:2010gn,Boehm:2010qt,Hutsi:2010ai,Lin:2010fba}
  have been devoted to extract indirect dark matter signatures or to
  set limits on a large class of WIMP models by investigating the
  Fermi-LAT diffuse gamma-ray measurements. Especially, a universal
  model-independent method can be used to place constraints on any
  dark matter model~\cite{Zhang:2009pr,Zhang:2009ut} through
  convolving {\it response functions} of signal-to-background with a
  specific injection energy spectrum of electrons and positrons.

In the WIMP cold dark matter scenario, the dark halos can form at very
high redshift, $z\approx60$, with a minimum mass of $\sim10^{-6}
M_\odot$~\cite{Green:2005fa,Loeb:2005pm,Bertschinger:2006nq,Profumo2006}
determined by the free-streaming limit and collisional damping leading
to a cutoff of the primordial power spectrum. This suggests that if
they can survive until the present day, an enormous number of dark
matter clumps (subhalos) are expected to be embedded in our
Galaxy. Recent numerical
  simulations~\cite{Diemand:2005vz,Gao:2007gh,Gao:2004au} confirm this
  prediction, although the role of the tidal effects from the baryonic
  component has yet to be quantified and there is no direct test by
observations. Nevertheless, several references have discussed
the resulting signals pointing out that subhalos can boost
annihilation rates and produce a distinct radial distribution of
emissions~\cite{Baltz:2004bb,Diemand:2006ik,Cline:2010ag,Pieri:2007ir,Kamionkowski:2010mi}.

Alternatively, a statistical analysis of the full-sky emission map  can be used to identify 
dark matter signatures, which can be promising if, e.g.,
one can identify features in the anisotropy power spectrum that 
characterize uniquely the dark matter spatial distribution with
respect to other astrophysical sources.
%given
%that the astrophysical emission profile closely follows the matter
%density as opposed to the squared dark matter density. 
In this respect, one is interested in quantifying the anisotropy
signatures from annihilating dark matter which can be
different from those of astrophysical origin particularly
at small scales.

The first calculations of this kind were performed analytically for
gamma-ray background anisotropies induced by both annihilating 
extragalactic dark matter halos~\cite{Ando2006,Ando:2006cr} as well as
Galactic subhalos~\cite{Ando:2009fp}.  Analytic calculations were also
performed for the anisotropic radio signatures from decaying dark
matter~\cite{Zhang:2008rs} and compared with the mean intensity and
angular power spectrum of the astrophysical and cosmological radio
background.

More recently anisotropies of the diffuse gamma-ray background from
annihilating dark matter in galactic subhalos have been computed using
a numerical approach in Ref.~\cite{SiegalGaskins:2008ge}.
  According to their results, Fermi-LAT should be sensitive enough to
  constrain the amount of dark matter substructure through the
  anisotropy signal discussed above.  The same numerical approach as
in~\cite{SiegalGaskins:2008ge} can be extended to the anisotropies
from both the galactic and extragalactic dark matter
contributions~\cite{Fornasa2009,Cuoco:2007sh,SiegalGaskins:2009ux,Hensley:2009gh,Zavala:2009zr}.
In general, however, both the mean flux and anisotropy signal are
contaminated by ordinary astrophysical sources including galactic and
extragalactic resolved point sources, structure formation
shocks~\cite{miniati03,Miniati:2007ke}, and normal
galaxies~\cite{pabr02,Ando:2009nk}.  Removal of such
contributions~\cite{Cuoco:2010jb}, can greatly improve the
detectability for searches of dark matter.

All of the above analyses were based on the gamma-rays from
direct annihilation into photons. 
However, the ICS radiation from energetic electron and positron pairs
produced by annihilation of dark matter also generates unique gamma-ray
signatures, in an energy range that depends on the dark matter model.
Thus in the present paper we study the gamma-ray anisotropies produced by
this ``secondary'' emission, which can be regarded as a new target for
indirect searches of signatures induced by Galactic substructure halos.
The important point is that the properties of the angular anisotropies
for emission from secondary electrons and positrons will be 
different in general than for prompt emission, because
the former will be affected by propagation effects in the Galaxy.
In fact, by adopting a spatial diffusion model typical for high
energy electrons and positrons in the Galaxy, we show that the angular power
spectrum is suppressed at small angular scales corresponding to the
distance diffusively traveled by the charged particles during their energy
loss time. As a result, for a typical dark matter model with mass
of 1 TeV and a canonical thermal freeze-out cross section
$3\times10^{-26} {\rm cm^3/s}$, the angular power spectrum from
inverse Compton scattering peaks at large angular scales. %{\brown (text removed)}
%, contrary to the gamma-ray angular power spectrum obtained from a simple analysis of the publicly available Fermi-LAT data.

The remainder of this paper is organized as follows. In
Sect.~\ref{sec:formulation}, we set up the formalism used in this
paper for calculating the mean intensity and angular power spectrum of
gamma-rays induced by dark matter annihilation and develop a robust
numerical scheme to simulate the small subhalos with masses down to
$\sim10^{-6} M_\odot$. We present the results of our calculations in
Sect.~\ref{sec:result}.  %{\brown (text removed)}
%and compare them with the Fermi observations. 
Finally, we summary our paper in
Sect.~\ref{sec:conclusion}. We will use natural units in which $c=1$
throughout.

\section{Formalism}\label{sec:formulation}
\subsection{Diffusion models}\label{subsec:diff}
The propagation of high energy electrons and positrons in the turbulent Galactic magnetic field can be described as a
diffusion-energy loss equation~\cite{schlickeiser02} for the electron-positron number density $n_e$, neglecting the convection and re-acceleration terms which
are only relevant for electrons and positrons below 10 GeV~\cite{Delahaye:2008ua}:
\begin{equation}
\frac{\partial n_e}{\partial t} = \nab \cdot \left( D(E,{\bf r}) \,\nab n_e \right) +  \frac{\partial}{\partial E} \left(b(E,{\bf r}) \, n_e \right) + Q(E,{\bf r})\,.
\label{eq:diffeqn}
\end{equation}
Here, the spatial diffusion coefficient is taken as $D(E,{\bf r}) =
D_0(E/\rm GeV)^\delta$; $b(E,{\bf r})$ is the energy loss term
and $Q(E,{\bf r})$ the source term, both of which are
described in more details below.
In the following we will often use the term
 electron as a short hand for both electrons and positrons. With the
assumption that spatial diffusion and energy loss coefficients are
spatially independent, for a given source distribution and boundary
conditions, the propagation equation can be solved analytically. We
assume the diffusion zone to be a cylinder with half-height $L$ of a few
kpc and radius $R\ga20\,$kpc. In this paper, we adopt the widely used
MED model which, compared to other models, i.e. the MIN and MAX models, predicts intermediate values
for the antiproton flux when fitted to reproduce the B/C ratio~\cite{Donato:2001ms} with $D_0=1.28\times10^{27}{\rm cm^2/s}$, $\delta=0.7$, and the half-height
of the diffusion zone $L= 4\,$kpc. The MIN and MAX propagation models would decrease and increase the predicted gamma-ray fluxes roughly by a factor of three and two, respectively. However, we verified that the shape of the angular power spectrum predicted by these three propagation models is basically the same. Since in the present paper we mainly focus on a new ``suppression'' feature in angular power spectrum of the gamma-ray component from ICS, we will use the MED model in the following calculations.

In this study, we investigate only the case in which electrons
  and positrons can escape freely at the boundaries of the diffusion
  zone. If the diffusion coefficient increases exponentially instead
  of the free-escape boundary condition, we expect that the angular
  power spectrum would be modified only at multipoles $l\ll10$ since
  the diffusion of electrons and positrons would flatten their distribution at length
  scales even larger than the whole diffusion zone. This implies that
  the angular power spectrum produced in the region of extremely large
  diffusion coefficient can not significantly influence the signature
  in the full-sky angular power spectrum we found at $l\sim10$. In
  other words, the anisotropies weakly depend on different boundary conditions.   

\subsubsection{Energy losses}\label{subsec:eloss}
At energies above 10 GeV, the dominant energy losses are synchrotron
radiation and inverse Compton scattering (ICS) on the interstellar
radiation field. Therefore, in the Thomson limit we write
$b(E,{\bf r}) = b_0\,E^2$ [for a more complicated treatment of energy
loss see Ref.\cite{Delahaye:2010ji}], where
$b_0=3\times10^{-16} \rm GeV/s$ for starlight (SL) , infrared (IR),
CMB photons and a magnetic field of $3\rm \mu G$. The Inter-Stellar
Radiation Field (ISRF) can be approximately characterized as a
superposition of three blackbody-like spectra with different
temperatures and normalization factors relative to a true
  black-body emitter: one for the CMB with $T_{CMB}=2.73\rm K$, for
the IR with $T_{IR} = 40.61\rm K$ and for the SL with $T_{SL}=3800 \rm
K$~\cite{Cirelli:2009vg}. 
The typical normalization of the SL and IR fields of radiation
depends on the position in the Galaxy. The averaged normalizations of
ISRF photon densities per energy used in this paper are
$8.9\times10^{-13}, 1.3\times10^{-5}$ and $1$ for SL, IR and CMB,
respectively~\cite{Cirelli:2009vg}. Although these normalizations are
valid only in the region with latitude $20^{\circ}>|b|>10^\circ$, we
have checked that changing the normalization of the ISRF affects the
gamma-ray emissions only weakly because an increased emission is
partly compensated by a decrease of the density of electrons and
positrons as energy losses increase. We also verified that the
deviation from the detailed numerical simulation with
Galprop~\cite{Strong:1998pw} is less than a factor of two for a
realistic spatial distribution of the ISRF~\cite{Porter:2006tb}.

\subsubsection{Dark matter model}\label{subsec:dm}
In the annihilating dark matter scenario, the source term can be written as 
\begin{equation}
  Q({\bf r},E)=\frac{1}{2}\left<\sigma v\right>
  \left(\frac{\rho({\bf r})}{m_\chi}\right)^2f_e(E) , 
\end{equation}
where $f_e(E)$ is the annihilation spectrum into electrons and
positrons at energy $E$. For simplicity, we assume mono-energetic
injection of the positron and electron in case of CP conservation,
namely $f_e = 2\delta(E-m_\chi)$. We choose $m_\chi = 1 \rm TeV$,
which can well fit the PAMELA excess~\cite{Cirelli:2008pk} while not
in conflict with gamma-ray observations by Fermi-LAT~\cite{Cirelli:2009vg}. Finally, we use $\left<\sigma
  v\right> =3\times 10^{-26} {\rm cm^3/s}$ to reproduce the correct relic
density for thermal freeze-out. By convolving our results for the gamma-ray spectra with the pair energy, our computational approach can be easily adapted to pair spectra
different from mono-energetic injection, such as from dark matter annihilating into $\mu^\pm, \tau^\pm$, and $W^\pm$.

% Since the astrophysical sources producing gamma-rays should highly
% contaminate the sky regions around the galactic center
% ($|b|<30^\circ$ for $|l|<40^\circ$) and the galactic plane, the
% signals in those regions are masked.

% Since the motivation of our paper is to explain the FERMI Haze which
% have a spherical morphology extended out to $20^\circ$ from the
% Galactic center, we averaged the ISRF on a rectangular region
% $'10\times60'$ yielding the normalizations which is
% $2.7\times10^{-12}, 7.0\times10^{-5}$ and $1$ for SL, IR and CMB,
% respectively~\cite{Cirelli:2009vg}.

\subsection{Green's function}\label{subsec:green}                 
As most of the electrons escape towards the $z-$direction, we impose
the Dirichlet boundary condition $n(x,y,z= |L|) = 0 $, at which the
particles can freely escape. We thus model the diffusion zone as an
infinite slab of half thickness $L$, and we take $L =4 \rm kpc$ for
the MED diffusion model.  The free-space Green's
function~\cite{Baltz:2004bb} for Eq.~(\ref{eq:diffeqn}) is
\begin{equation}
G_{\rm free}\left[{\bf r},{\bf r}^\prime,\lambda_D(E,E')\right]= \frac{1}{b(E)}\frac{1}{(\pi\lambda_D^2)^{3/2}}\e^{-({\bf r}-{\bf r}^\prime)^2/\lambda_D^2}\,,
\end{equation}
where we have defined the diffusion length as
\begin{equation}
\lambda_D^2(E,E') \equiv 4\int_E^{E'} \frac{D(E)}{b(E)} dE = 4D_0\frac{\rm GeV}{b_0}\left(\frac{(E/{\rm GeV})^{\delta-1} -(E'/{\rm GeV})^{\delta-1}}{1-\delta}   \right)\,,
\label {eq:d}
\end{equation}
 which is the average distance $e^+e^-$ diffuse through during their energy loss time.
Then, the Green's function satisfying appropriate boundary conditions
can be obtained by considering a
series of image charges at positions $x_i=x, y_i=y, z_i = (-1)^iz +
2i\cdot L $. One can verify that
\begin{equation}
G_{2L}({\bf r},{\bf r}^\prime,\lambda) = \sum^\infty_{i=-\infty} (-1)^iG_{\rm free}
({\bf r},{\bf r}^\prime_i,\lambda)
\end{equation}
fulfills the Dirichlet boundary condition. Thus, the general solution to Eq.~(\ref{eq:diffeqn}), in the limit of time-independent sources and electron/positron 
number densities which already reached equilibrium, is given by 
\begin{equation}
n_e({\bf r},E) = \frac{1}{b(E)}\int d^3{\bf r}^\prime\int_E^\infty 
dE^\prime G_{2L}({\bf r}-{\bf r}^\prime, \lambda_D(E,E^\prime))
Q({\bf r}^\prime,E^\prime)\,.
\end{equation}

For the diffuse gamma-ray emission we are more interested in the column density of electrons,
 \begin{equation}\label{eq:column}
\sigma_e(l,b,E) = \int_0^{l_{max}} d\ell~ n({\bf r},E)\,,
\end{equation}
than in the local space density of electrons.
The observer is located at the solar system. In galactic coordinates,
a point in cartesian coordinates $(x,y,z)$ at a distance $\ell$ from the observer
is given by
\begin{eqnarray}
x&=&\ell\,\cos b\,\cos l,\\
y&=&\ell\,\cos b\,\sin l,\\
z&=&\ell\,\sin b\,,
\end{eqnarray}  
where  $l$   and  $b$  are   the  galactic  longitude   and  latitude,
respectively. We truncate the integral in Eq.~(\ref{eq:column}) at the
edge  of   the  diffusion  zone, beyond  which  particles   are  not
confined, $z_{max}=L$ or $\ell_{max}  = L/|\sin b|$. The line-of-sight integral
can directly act on the free Green's function~\cite{Baltz:2004bb},
\begin{eqnarray}
G_{\rm free}^\sigma(l,b,{\bf r}^\prime,\lambda_D)&=&
\int_0^{\ell_{\rm max}} d\ell\,G_{\rm free}(\ell{\bf n},{\bf r}^\prime,\lambda_D)=\\
&=&\frac{e^{[({\bf n}\cdot{\bf r}^\prime)^2-({\bf r}^\prime)^2]/\lambda_D}}{2\pi \lambda_D^2b(E)}
\left[\erf\left(\frac{\ell_{\rm max}-{\bf n}\cdot{\bf r}^\prime}{\lambda_D}\right)-
\erf\left(\frac{-{\bf n}\cdot{\bf r}^\prime}{\lambda_D}\right)\right]\,,\nonumber
\end{eqnarray}
where we have defined the unit-vector ${\bf n}\equiv{\bf r}/\ell$. The Green's function satisfying the boundary condition is thus
\begin{equation}
G_{2L}^\sigma(l,b,{\bf r}^\prime,\lambda) = \sum^\infty_{i=-\infty} (-1)^iG_{\rm free}^\sigma(l,b,{\bf r}^\prime_i,\lambda)\,.
\end{equation}
The column density of electrons therefore reads
\begin{equation}\label{eq:sigma}
\sigma_e (l,b,E) = \frac{1}{b(E)}\int d^3{\bf r}^\prime\int_E^\infty dE^\prime~ G_{2L}^{\sigma}\left[l,b,{\bf r}^\prime,\lambda_D(E,E')\right]
Q({\bf r}^\prime,E^\prime)\,.
\end{equation}
In the limit of $\lambda_D \gg r_s$, where $r_s$ is the scale radius
of the subhalo profile, the subhalo can be regarded as a point-like
source. Eq.(~\ref{eq:sigma}) can then be simplified to
\begin{equation}\label{eq:ssigma}
\sigma_e (l,b,E) = \sum _k \frac{1}{b(E)}\int_E^\infty dE^{\prime}~G_{2L}^{\sigma}
\left[l,b,{\bf r}_k,\lambda(E,E^\prime)\right]j_{k}({\bf r}_k),
\end{equation}
where $j_k(E)=\int d^3{\bf r}~Q_k({\bf r},E)$ for a given subhalo
source $Q_k$ located at ${\bf r}_k$. For the largest subhalos with masses larger than $\simeq10^{9} M_\odot$, their radius $r_s$ can be somewhat larger than the diffusion length $\lambda_D$. Nevertheless, their contribution to the total flux is a factor $\sim10^{4}$ smaller than the flux from the smaller subhalos. Therefore, neither the mean intensity nor the dimensional angular power spectrum (see Sect.~\ref{subsubsec:cl}) relies significantly on the distribution of electrons in the largest subhalos. As a result, for our purposes
we can apply Eq.~(\ref{eq:ssigma}) to all subhalos even if $r_s>\lambda_D$.

\subsection{Halo function}\label{subsec:halo}
Alternatively, for primary electrons and positrons from the smooth host dark matter halo, the Bessel-Fourier scheme~\cite{Delahaye:2007fr,Maurin:2001sj} can require less computational time than the Green's function. The electron and positron number density after propagation can be expressed as
\begin{equation}
n_e(r,z,E) =  \frac{1}{b(E)} \int_E^{M_\chi}dE^\prime~f_e(E')~I(r,z,E,E^\prime),
\end{equation}
and $I(r,z,E,E^\prime)$ is the halo function defined by
\begin{equation}
I(r,z,E,E^\prime) = \sum_i\sum_n J_0\left(\frac{\alpha_ir}{R} \right)\sin\left[\frac{n\pi(z+L)}{2L}\right]e^{-\left[\left(\frac{n\pi}{2L}\right)^2+\frac{\alpha_i^2}{R^2}\right]\frac{\lambda_D^2}{4}} R_{i,n}\,.
\end{equation}
Here, the $\alpha_i$ are the zeros of the Bessel function $J_0$ and $R_{i,n}$ are the coefficients of the Bessel-Fourier transform of the source term. 

\subsection{Diffuse Emission: Inverse Compton spectrum}\label{subsec:IC}
For relativistic electrons and positrons with energy $E$ up scattering
background photons from energy $\epsilon$ to $E_\gamma$, the
emitted inverse Compton power per energy interval is
\begin{equation} 
P_{\rm IC}(E_\gamma,E) = E_{\gamma}\int~d\epsilon~n(\epsilon)
\frac{d\sigma}{dE_\gamma}(E_\gamma,\epsilon,E)\,,
\end{equation}
where $n(\epsilon)$ is the differential ISRF photon number density, while the differential cross section $(d\sigma/dE_\gamma)(E_\gamma,\epsilon,E)$ is given by the Klein-Nishina formula~\cite{Rybicki}.
Folding $P_{\rm IC}$ with the spectral distribution of the equilibrium number density of electrons and positrons, we get the emissivity of IC photons of energy $E_\gamma$,
\begin{equation} 
j_{\rm IC}(E_\gamma) = \int~dE~n_e(E)~P_{\rm IC}(E_\gamma,E),
\end{equation}
which yields the IC intensity at energy $E_\gamma$ by the line-of-sight integral
\begin{equation}
I_{\rm IC}(E_\gamma) = \frac{1}{4\pi}\int~d\ell~j_{\rm IC}(E_\gamma,{\bf r})\,. 
\end{equation}

According to the Eq.~(\ref{eq:sigma}), the IC intensity from the electrons and positrons can be simplified to
\begin{equation} \label{eq:ICI}
I_{\rm IC}(l,b,E_\gamma) = \frac{1}{4\pi}\int~dE~P_{\rm IC}(E_\gamma,E)~\sigma_e(l,b,E)\,.
\end{equation}
There is a well known ``delta-function approximation'' where an
electron with energy $E$ inverse Compton scattering black-body photons
with temperature $T$ emits photons with a characteristic energy
$E_\gamma$ in both the Thomson and extreme Klein-Nishina
limits~\cite{Petruk:2008bb}, $\ie$, $P_{\rm
    IC}(E_\gamma,E)=P_{\rm IC}(E)\delta[E_\gamma-E_c(E)]$, where
  $P_{\rm IC}(E)$ is the total IC energy loss rate of the
  electron. The numerical calculations show that $E_c(E)$ may be
  approximated by $E_c(E)\simeq4k_BT(E/m_e)^2$ and in the Thomson
  regime one has $P_{\rm IC}(E)=(16e^4\pi/3)u_b E^2/m_e^4$, where $e$
  is the electron charge and $u_b$ is the background photon energy
  density. Eq.~(\ref{eq:ICI}) can thus be simplified to
\begin{equation} \label{eq:simICI}
I_{\rm IC}(l,b,E_\gamma) = \frac{1}{16\pi}\,\sigma_e(l,b,E)\frac{m_eP_{\rm IC}(E)}{\sqrt{E_{\gamma}k_BT}}\,,
\end{equation}
where the electron/positron energy $E$ is related to the gamma-ray
energy  $E_\gamma$ through the following condition:
$E_\gamma=E_c(E)$.  This relation reproduces the known slope
$I_{\rm IC}(E_\gamma)\propto E_\gamma^{-(s-1)/2}$ for an electron spectrum
of $\sigma_e(E) \propto E^{-s}$. In the Thomson limit and 
for our choice of monoenergetic injection of electron-positron pairs, the index
$s\simeq2$ in the stationary situation if the energy loss term
dominates on the right hand side of Eq.~(\ref{eq:diffeqn}), as is
usually the case for electron energies above 10 GeV. For $E\la\,$TeV
we are always in the Thomson limit.
  
\subsection{Galactic halo substructure}\label{subsec:subst}  
In the present section we discuss contributions to the source term
$Q({\bf r},E)$ both from the smooth host halo and the individual
subhalos. The two most important features of the subhalos are their
mass distribution and their spatial distribution. In the following, we
briefly summarize the properties of subhalos used in our calculations
for which we adopt the same description as in
Ref.~\cite{SiegalGaskins:2008ge}, and references therein.

\subsubsection{Subhalo radial distribution}
There are two widely used scenarios for describing the subhalo radial distribution. One is unbiased relative to the Galactic smooth component (host halo) with the NFW density profile~\cite{NFW} given by 
\begin{equation}
\rho_{\rm NFW}(r) = \frac{\rho_s}{x(1+x)^2}\,,
\end{equation}
with $x\equiv r/r_s$, where $r_s$ is a scale radius and $\rho_s$ is the characteristic density. For the case where the subhalo distribution is anti-biased compared to the smooth component, we use the fitting formula of the subhalo radial distribution from Gao {\it et al}~\cite{Gao:2007gh,Gao:2004au}. The cumulative fraction of subhalos within a given radius is
\begin{equation}
\frac{N_{anti}(<\zeta)}{N_{tot}} = \frac{(1+a\,c_{200})\,\zeta^\beta}{1+a\,c_{200}\,\zeta^\gamma}\,,  
\end{equation}   
with $\zeta\equiv r/r_{200}$, $a=0.244$, $\beta=2.75$, $\gamma=2$ and $c_{200}\equiv r_{200}/r_s$ is the host halo concentration. Furthermore, $N_{tot}$ is the total number of subhalos within virial radius $r_{200}$ of the host halo.
For a host halo with the NFW density profile we adopted $\rho_s=0.2~{\rm GeV/cm^3}$,
$r_s=21.7\,$kpc and $c_{200}=12$~\cite{Fornengo:2004kj}.
 
\subsubsection{The subhalo mass function}\label{subsubsec:dmmass}
Recent simulations suggest that the cumulative number of subhalos above a given
mass $M$ in units of the solar mass  can be fitted simply by a power law~\cite{Diemand:2007qr},
\begin{equation}\label{eq:dndm}
N(>M)\simeq64\left(\frac{M} {10^{8}M_\odot}\right)^{-\alpha_m}\,.
\end{equation}   
In the present paper, we simulate the subhalos with mass down to around one Earth mass, $M_{\rm min} = 10^{-6}M_\odot$ and choose $\alpha_m = 0.9$, as suggested by simulations. We find that choosing $\alpha_m$ between $0.8$ and $1$~\cite{SiegalGaskins:2008ge} would change the mean intensity by roughly a factor of $20$. 

\subsubsection{The subhalo density profile}
More recent simulations suggest that the central structure of dark
matter halos deviates from the NFW profile in the innermost
regions, which can be well reproduced by an Einasto
density profile.
%\blue{Einasto proposed that a density profile whose logarithm
%  is a power law of the radius can well fit the simulation data.} 
Therefore, here we use the Einasto profile with the parameter $\alpha
=0.16$ given by Gao {\it et al}~\cite{Gao:2007gh},
\begin{equation}
\rho(r) = \rho_s\exp\left(\frac{2}{\alpha}\right)\exp\left[-\frac{2}{\alpha}\left(\frac{r}{r_s}\right)^\alpha\right]\,.
\end{equation}
The relation of halo concentration and mass is given by 
\begin{equation}\label{eq:c}
c(M) = 398.1 \left(\frac{M}{M_\odot}\right)^{-0.138}\,.
\end{equation}
The slope and normalization are consistent with those found by Bullock {\it et al}~\cite{Bullock:1999he}.

\subsection{The angular power spectrum}\label{subsubsec:cl}

The angular power spectrum of the emission maps can be calculated by using the public HEALPix package~\cite{Gorski2005}. We define the dimensionless quantity $\delta I(\psi,E_\gamma) \equiv (I(\psi,E_\gamma) - \langle I \rangle)/\langle I\rangle$ as a function on the sphere, which can be expanded in spherical harmonics $Y_{lm}$ as
\begin{equation}
\delta I = \sum_{l=0}^{l_{max}}\sum_{m=-l}^{l} a_{lm}Y_{lm}(\psi)\,,
\end{equation}
where I($\psi$) describes the gamma-ray intensity in the direction $\psi$. The {\it dimensionless} angular power spectrum of $\delta I$ is given by the coefficients
\begin{equation}
C_l = \frac{1}{2l+1} \sum_{m=-l}^{l} |a_{lm}|^2\,.
\end{equation}
Note that the measured dimensional amplitude $C_l^I$ of the total angular power spectrum in units of intensity squared can be obtained by multiplying the dimensionless angular power spectrum of a given component, $C_{l,i}$,
with its mean intensity squared, $\langle I_i \rangle^2$ and summing over the components,
\begin{equation}\label{C_l}
  C_l^I=\sum_i\langle I_i \rangle^2C_{l,i}\,,
\end{equation}
where in our case the sum basically runs over the contributions of the smooth host halo, the subhalo distribution and other astrophysical foregrounds and backgrounds. Here we assumed
that the different contributions are uncorrelated.

\subsection{Numerical scheme}
In numerical calculations, taking into account subhalos down to masses
$\sim10^{-6} M_\odot$ in the angular power spectrum would require the
generation of $\sim10^{16}$ subhalos within the diffusion zone in a
given Monte Carlo realization. This is not very practical and
  severely limits the number of realizations one can simulate. Thus,
we need to develop a sound scheme to circumvent these computational
limits. Here we take advantage of a few simple facts:
(i) Two terms  contribute to the fluctuation anisotropies in units of average intensity squared;
the one-subhalo term ($C^I_{1h}$)  which is Poissonian noise and the two-subhalo term ($C^{I}_{2h}$)
due to the cross-correlation of substructure arising from
their radial distribution within the host halo (see Eqs.~$(19)$ and $(20)$ of Ref.~\cite{Ando:2009fp})).
(ii) Qualitatively speaking, the two-subhalo term scales linearly with the square of the average flux whereas the one-subhalo term  is proportional to an integral of squared luminosities of subhalos.  Since subhalo luminosity $L_{sub}$ is related to its mass $M$ via $L_{sub}\propto \int dV_{sh} \rho_{sh}^2(r,M) \approx Mc(M)^3$, by using Eq.~(\ref{eq:c}), the correlation between gamma-ray luminosity and mass thus is positive, i.e., $L_{sub}\propto M^{0.6}$. One-subhalo term and two-subhalo term per decade of subhalo mass then can be estimated as $C^I_{1h} \propto dN/d\log(M) L_{sub}^2 \approx M^{0.3}$ and $C^I_{2h} \propto (dN/d\log(M) L_{sub} )^2\approx M^{-0.6}$. Therefore, large and relatively rare substructures dominate 
 the one-subhalo term, but negligibly to the two-subhalo term, and the small but numerous substructure halos dominate the two-subhalo term. 

%(iii) At large angular scales, corresponding to the small $l$, the angular power spectrum usually is dominated by the two-subhalo term whereas the one-subhalo term become more important compared with two-subhalo term at smaller angular scales}. 

%because their
%Poissonian noise is suppressed as \blue{$N(>M)^{-1/2}$};}

%(i) the anisotropies are Poissonian in nature; (ii) thus
%the small but numerous substructure halos dominate
%the mean flux but contribute negligibly to the fluctuation anisotropies
%which are suppressed as $N(M)^{-1/2}$; (iii) 
%the fluctutation anisotropies are dominated by the large and relatively
%rare substructures.
% \red{We take advantage of the fact that the gamma-ray
%   intensity is dominated by the smallest subhalos, whereas
%  the fluctuations
%   of the gamma-ray fluxes  come from the largest subhalos depend on
%   different realizations due to the rare number of those brightest
%   sources}.
There is thus a dividing mass of substructures, $M_0$, below which their
contribution to one-subhalo term  are negligible. The numerical value of this
mass depends on the assumed radial distribution of the substructure
halos (see below). In any case, it turns out that halos with
$M<M_0$ have a characteristic radius, $r_s$, much smaller 
than the diffusion
length of the gamma-ray emitting electrons, i.e. $\lambda_D\gg r_s$.
Thus, effectively all halos below $M_0$ have the same
image on the sky in ICS with a characteristic size given by $\lambda_D$.
%This further
%simplifies our calculations because besides minor poissonian noise,
%the {\it dimensionless anisotropy power spectrum} from subhalos
%with $M<M_0$ is the same for each mass decade. 
This further simplifies our calculations because besides minor poissonian
 noise, the angular power spectrum
 from subhalos with $M<M_0$ is  simply
%just given by the emission profile
 obtained by convoluting the subhalo spatial distribution with the
 mass-independent ICS subhalo image.
% of a subhalo which is mass-independent for $M<M_0$.
Thus the contribution
from subhalos with $M<M_0$ is taken into account straightforwardly 
 as follows:
We generate a distribution of subhalos with mass in the decade 
$10^{-6}- 10^{-5} M_\odot$ to obtain the full-sky map. 

The number
of generated subhalos is typically much smaller than the actual value,
$N(10^{-6}<M/M_\odot<10^{-5})$, but is sufficiently large  that the two-subhalo term divided by the mean intensity squared has converged.
The intensity of the map and the corresponding two-subhalo term thus can be rescaled to the value appropriate for $N(10^{-6}<M/M_\odot<M_0)$ halos. In this way, the simulated two-subhalo term of the angular power spectrum can fairly approach the actual one. 

%The map contributed from halos
%in all other mass decades up to $M_0$ can be obtained by simply rescaling
%the intensity appropriate for halos in that mass decade. 

%(Alternatively, the calculation can be 
%repeated for halos in all other mass decades up to $M_0$.)

Since the one-subhalo term is not linearly proportional to the square of the average intensity and is contributed mostly by larger subhalos, we thus complete the calculation by adding the contribution of subhalos
with mass $M>M_0$, which can now be done with a direct Monte Carlo
simulation. 
%
% We thus first compute the contribution of large subhalos
% down to a mass scale $M_0$ {\brown using a suitable number of direct 
% realizations}, where $M_0$ is a limiting mass to be specified below.
% We then generate a
% computationally tractable number of small subhalos in the mass range
% $10^{-6}- 10^{-5} M_\odot$ to obtain the full-sky map, while ensuring
% that the induced angular power spectrum is not changed when increasing
% the number of subhalos in this mass range. 
% Finally, we normalize the
% resulting sky map to the correct mean intensity \blue{from subhalos
%   with masses from} $10^{-6}M_\odot$ to $M_0$ which can be calculated
% analytically. Based on the fact that the diffusion length
% $\lambda_D\gg r_s$, the characteristic size of subhalos, the
% fluctuations, therefore, almost exclusively depend on the same
% diffusion length for all subhalos provided their number is
% sufficiently large. Thus the {\it \red{dimensionless~} anisotropy
%   power} from small subhalos are the same for each mass decade. That
% is why we can linearly normalize the map without introducing spurious
% features in the angular spectrum.  Finally, we combine the directly
% simulated full-sky map with the normalized one. 
%
Specifically, we find that our calculation of the angular power spectrum
reaches convergence when we use at least $10^5$ subhalos for each mass
decade from $10^{-6} M_\odot$ to $M_0$ within the diffusion zone,
when we use $M_0 = 10^{4} M_\odot$ for the unbiased radial
distribution and $M_0=10^{2} M_\odot$ for the anti-biased case. 

Note that this method can not be applied to the angular power
spectrum of the direct annihilation component since in this case the
profile of the emitting region depends on the profile of each subhalo
rather than on the identical diffusion length which just depends on
the energy of electrons and positrons for all subhalo masses.

\section{Results for Galactic Dark Matter }\label{sec:result}
In this section, we make use of the formalism derived in the previous
section to compute the gamma-ray mean intensity map and associated
anisotropy power spectrum due to both the smooth host halo component
and the substructure halos.  We also study how the map morphology and
anisotropies depend on the radial distribution of subhalos and of the
gamma-ray energy. 
% Finally, we discuss the detectability by Fermi-LAT.
%In this section I will show the preliminary results of full-sky emission map and the corresponding power spectrum, while comparing the IC case with the direct one to illustrate the diffusion effect on both emission pattern and anisotropy. 

\begin{figure}[t]
\begin{center}
\epsfig{file=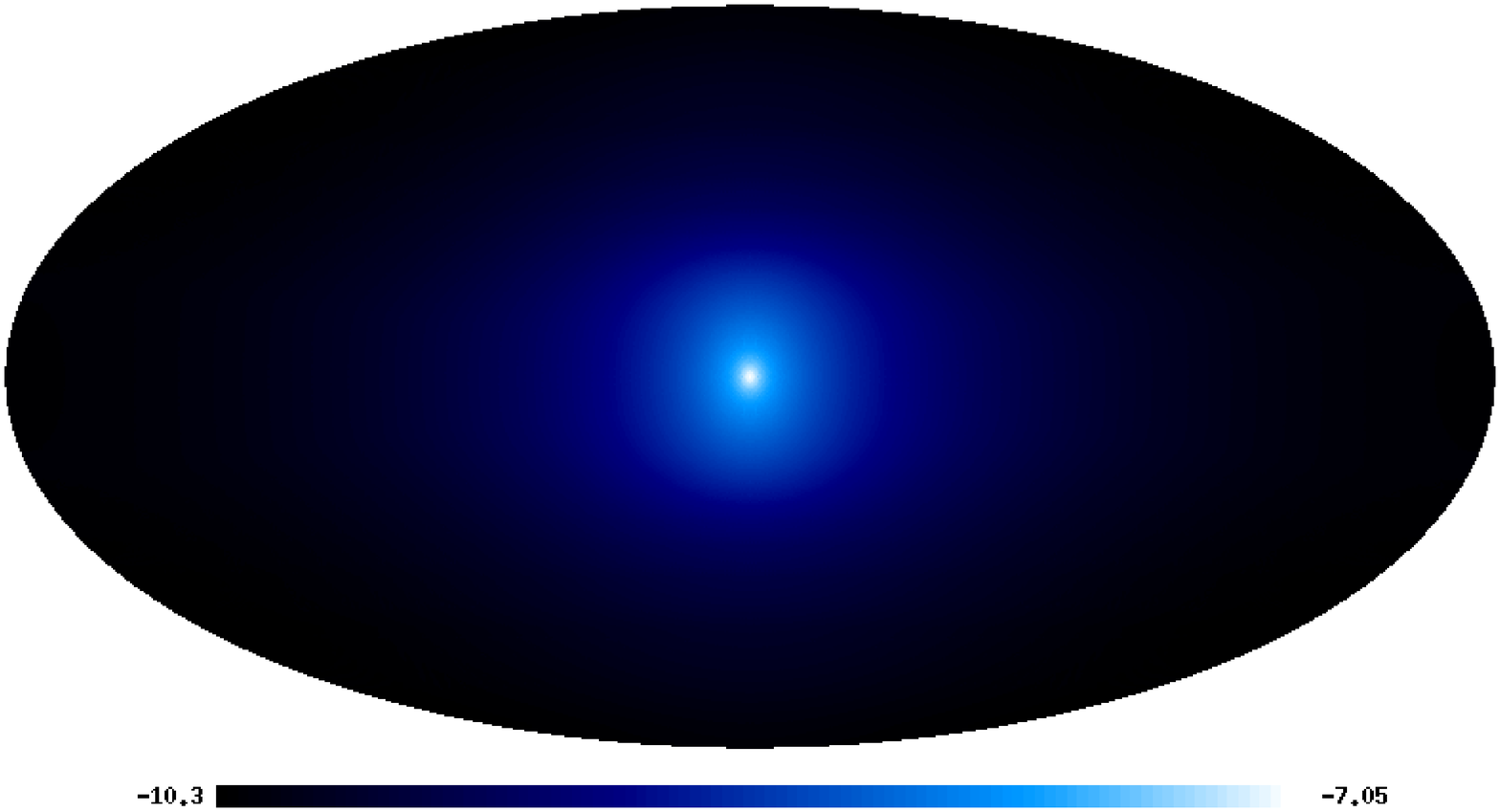, height=4.2cm}
\epsfig{file=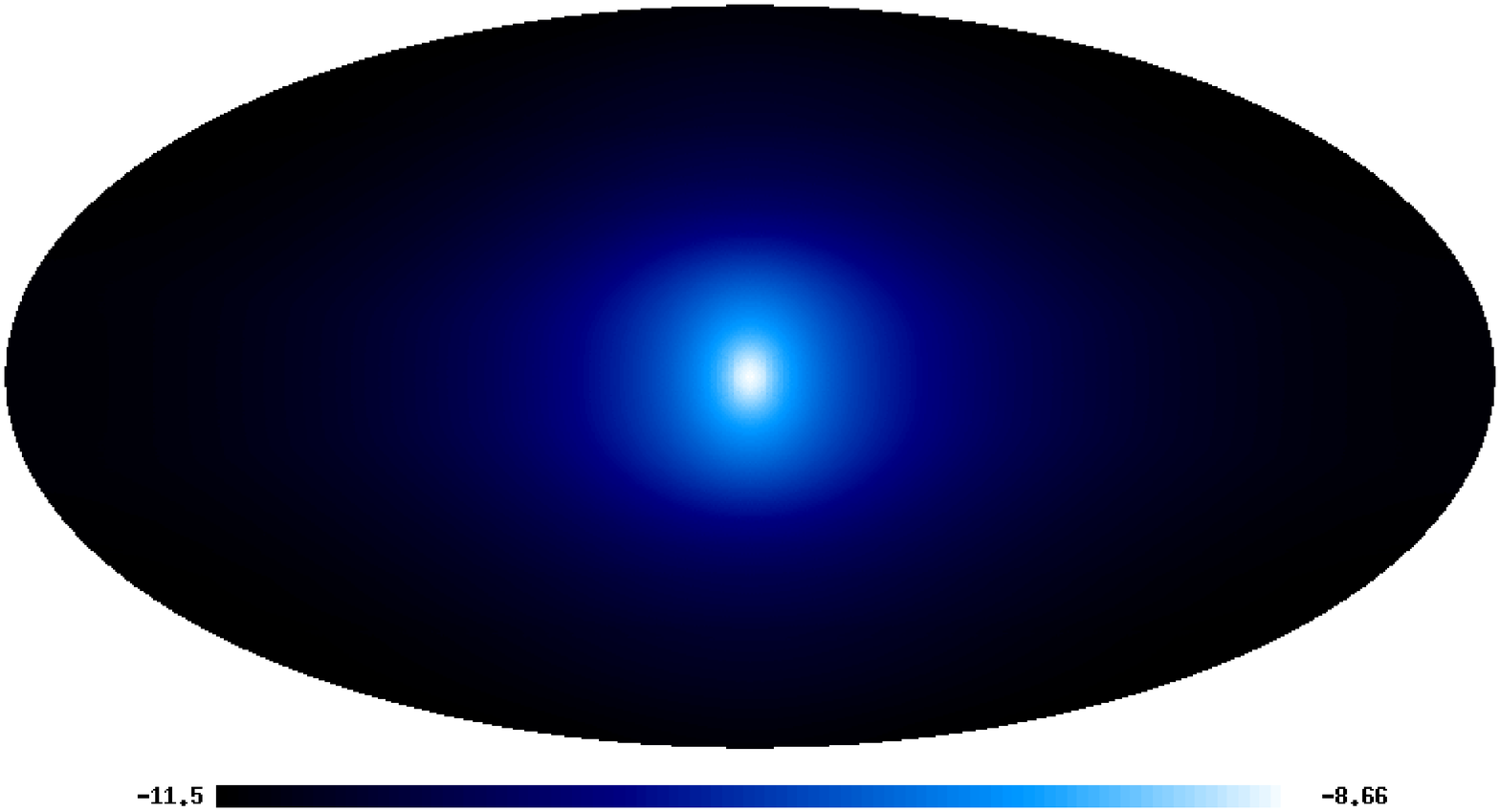, height=4.2cm}
\epsfig{file=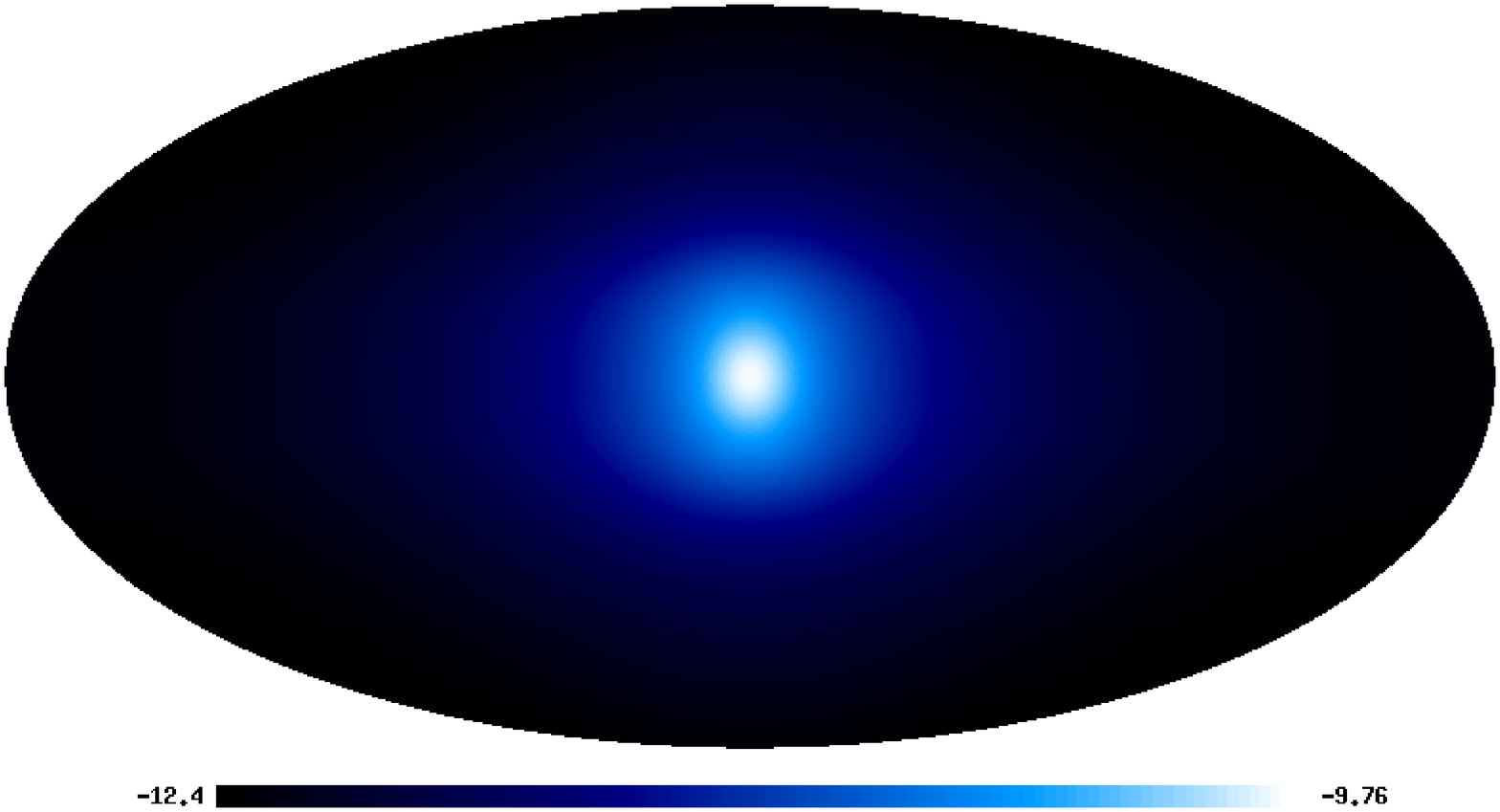, height=4.2cm}
\end{center}
\caption{\label{fig:diff-host} Sky maps of gamma-ray emission in Galactic coordinates
at 1 GeV (top-left), 10 GeV (top-right) and 100 GeV (bottom) due to ICS from the host halo. The color scaling is logarithmic, and the unit is  $\rm 1/s/cm^2/sr$.}
\end{figure}

\subsection{Diffuse Gamma-ray Emission}\label{subsec:dg}
We first consider the gamma-ray emission produced by dark 
  matter annihilating into $e^+e^-$ pairs scattering off the ISRF both
  in the smooth host halo and the subhalos. Using the formalism
developed in Sect.~\ref{sec:formulation}, the fluxes from the smooth
host halo at 1 GeV, 10 GeV and 100 GeV are shown in
Fig.~\ref{fig:diff-host}. Here we use galactic coordinates, where an
observer is located at 8.5 kpc from the Galactic center. We adopt the
NFW density profile with the parameters given in
Sect.~\ref{subsec:subst}, and find, as expected, that most of the
signal comes from the central Galactic region where the dark matter is
highly concentrated. The values of the mean gamma-ray intensity from
both the host halo and the substructure are also reported in
Tab.~\ref{tab:I} for the above three photon energies. We find that in
both cases the mean gamma-ray intensity decreases faster than $\sim
E_\gamma^{-1/2}$ in Eq.~(\ref{eq:simICI}). This is because there will
be a high energy cut-off in $E_\gamma$ once the  required
energy of the parent
pairs $E\simeq m_e[E_\gamma/(4k_BT)]^{1/2}$ scattering off the low
energy background photons exceeds the maximum energy $M_\chi$ produced
by dark matter annihilation. The effect is even stronger when 
the delta-function approximation for ICS is assumed, as in our case. 
As a consequence,
all background photon energies at CMB energies and above contribute to
the 1 GeV ICS photon flux, whereas only the IR and SL contribute to
the 10 GeV photon flux and only the SL contributes to the 100 GeV
photon flux, as also summarized in Tab.~\ref{tab:E_e}. Furthermore,
using the relation between $E$ and $E_\gamma$ and the expression for
$P_{\rm IC}(E)$, for an electron spectrum $\sigma_e\propto E^{-s}$
Eq.~(\ref{eq:simICI}) gives the scaling $I_{\rm IC}(E_\gamma)\propto
u_bE_\gamma^{(1-s)/2} T^{(s-3)/2}$. Using the different normalizations
for the CMB, IR and SL densities, then would suggest that the gamma-ray
intensity originating from $e^+e^-$ pairs scattering off CMB photons
is about 5 times larger than the one from scattering off the IR
photons and about 10 times lager than the contribution from scattering
off SL photons, provided there is no restriction from the
kinematics. However, the $e^+e^-$ pairs produced by annihilating dark
matter do  have an absolute cutoff at the parent particle energy. As a
result, the more detailed numerical results show that most of the
gamma-ray intensity at 1 GeV is produced by pairs with
$E\simeq526~{\rm GeV}$ scattering off the CMB, whereas pairs with
$E\simeq431 ~{\rm GeV}$ scattering off IR photons dominate the 10 GeV
gamma-ray flux and pairs with $E\simeq141 ~{\rm GeV}$ scattering off
SL dominate ICS photons at 100 GeV.

\begin{figure}[t]
\begin{center}
\epsfig{file=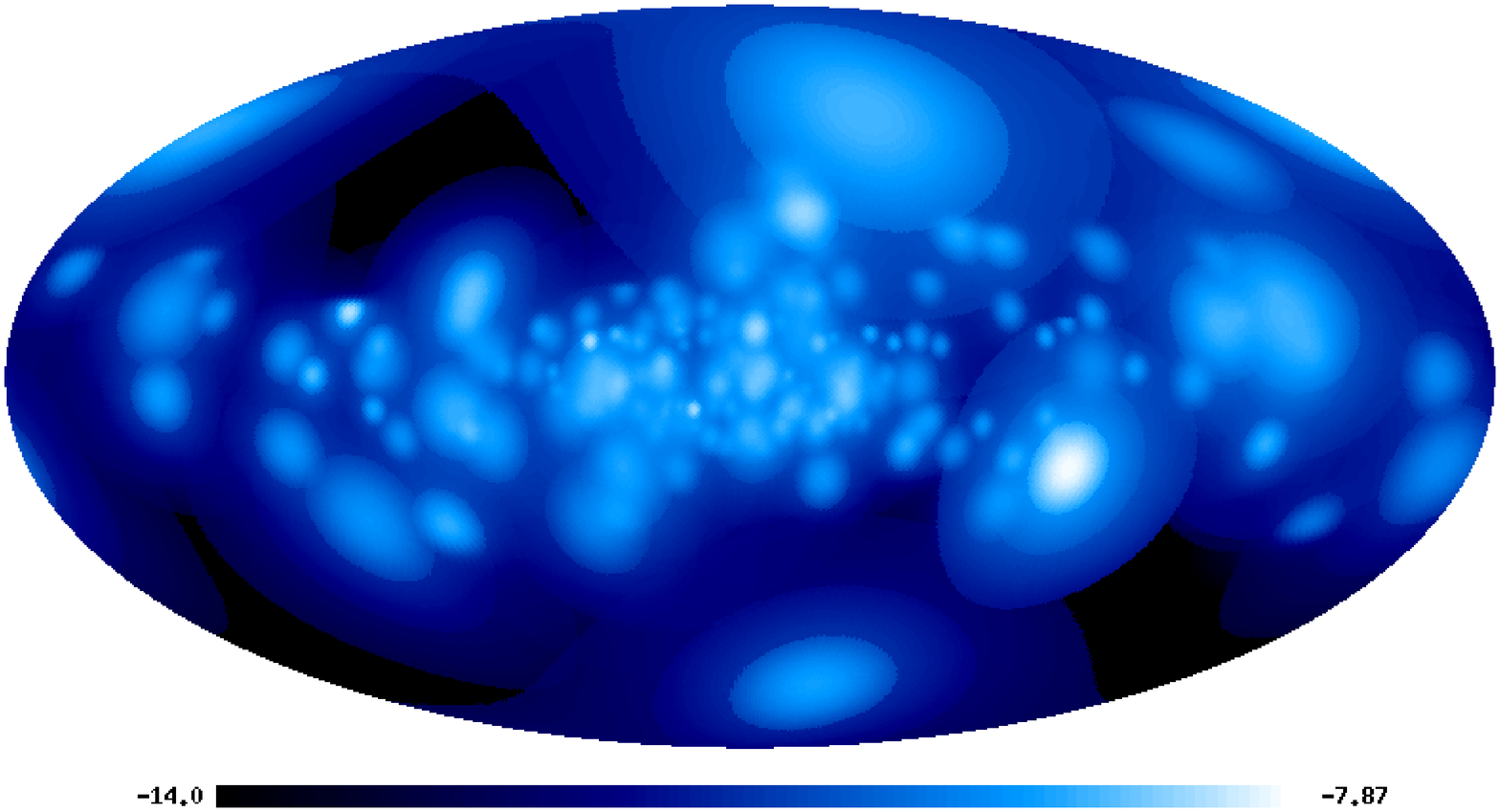, height=4.2cm}
\epsfig{file=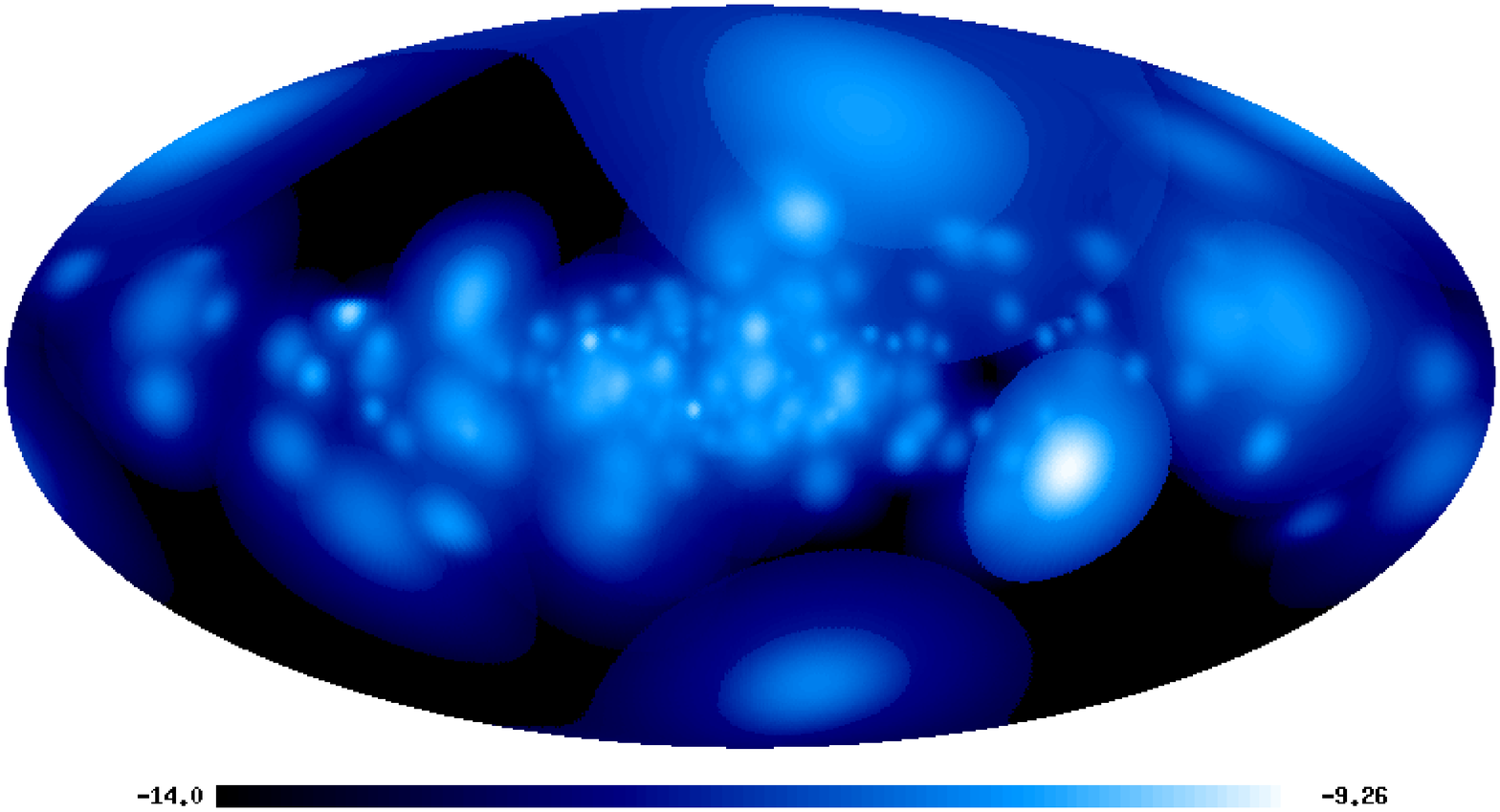, height=4.2cm}
\epsfig{file=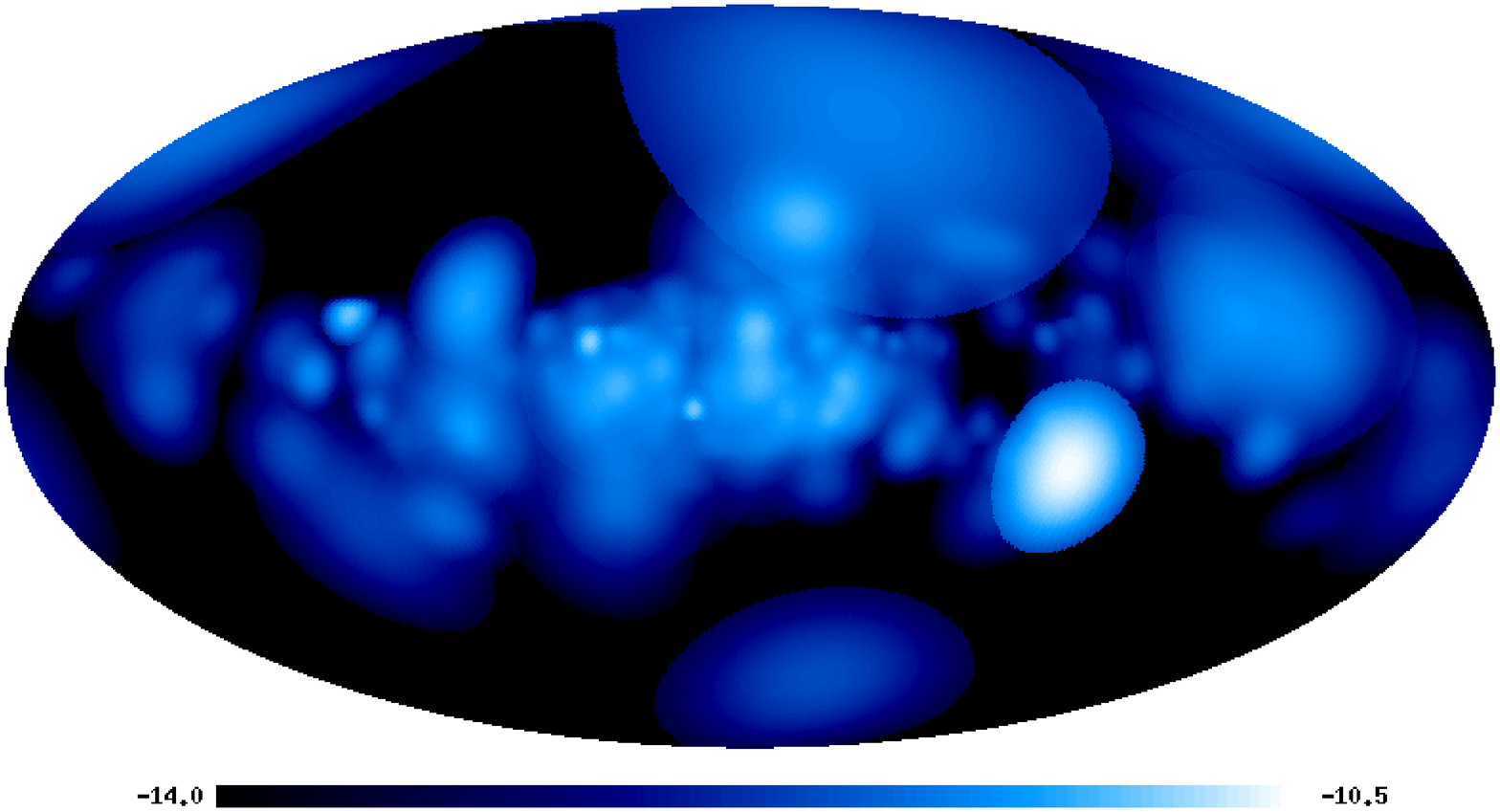, height=4.2cm}
\end{center}
\caption{\label{fig:diff-unbias}Sky maps of gamma-ray emission in Galactic coordinates at 1 GeV (top-left), 10 GeV (top-right) and 100 GeV (bottom) due to ICS from one realization of subhalos for the unbiased radial distribution and a minimum subhalo mass of $10^{6}M_\odot$. The color scaling is logarithmic, and the unit is  $\rm 1/s/cm^2/sr$.}
\end{figure}

\begin{figure}
\begin{center}
\epsfig{file=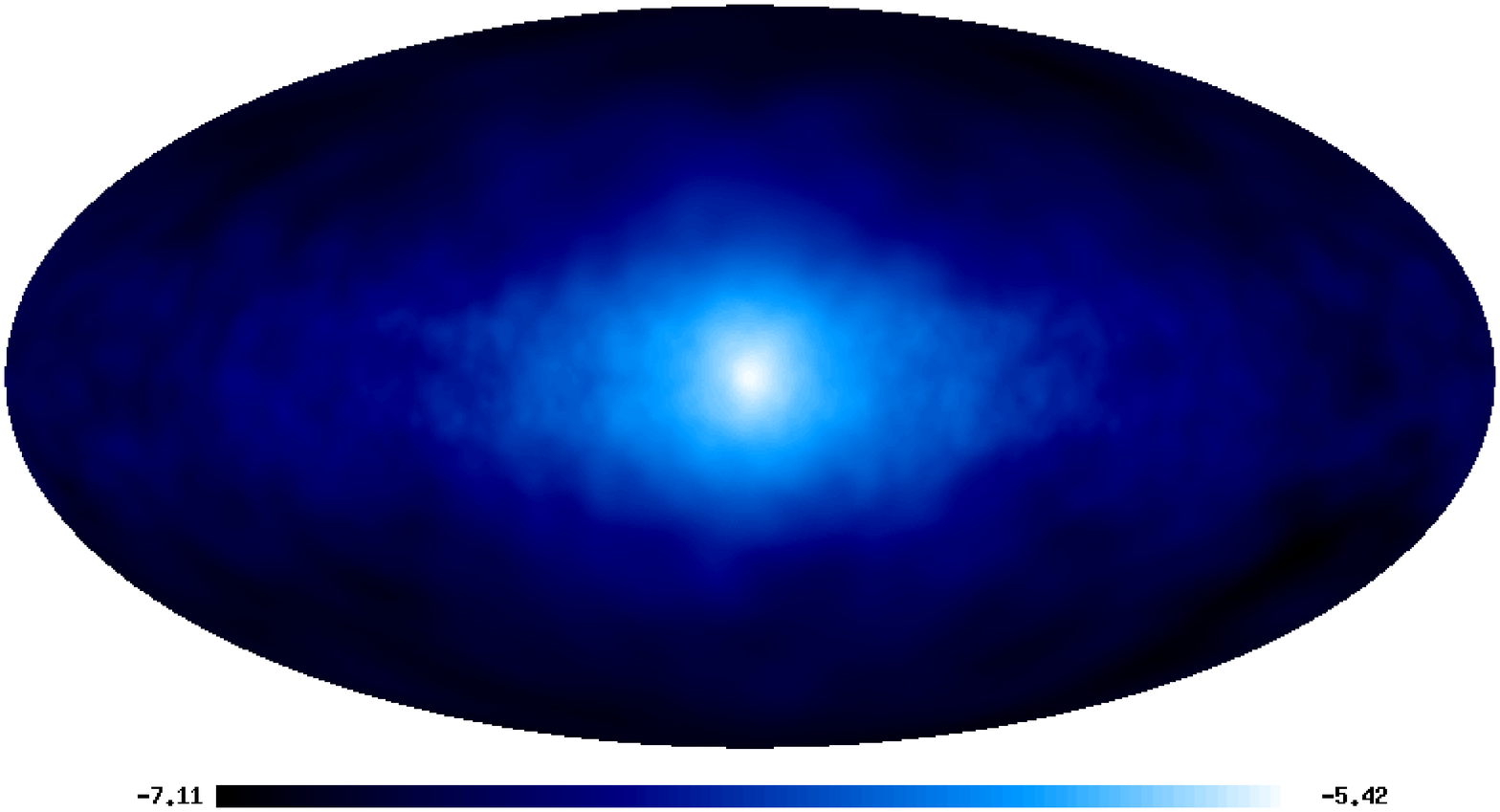, height=4.2cm}
\epsfig{file=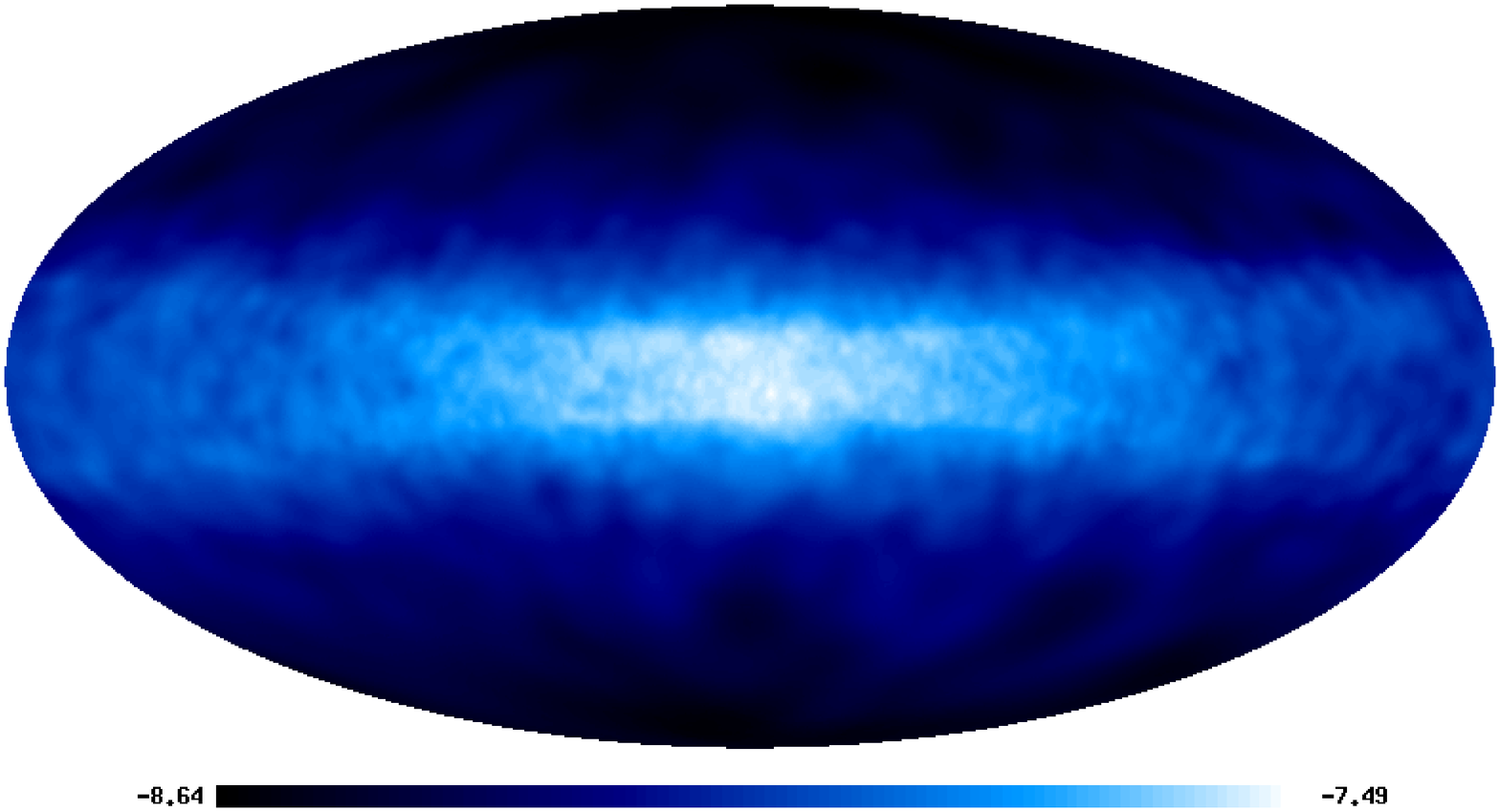,height=4.2cm}
\epsfig{file=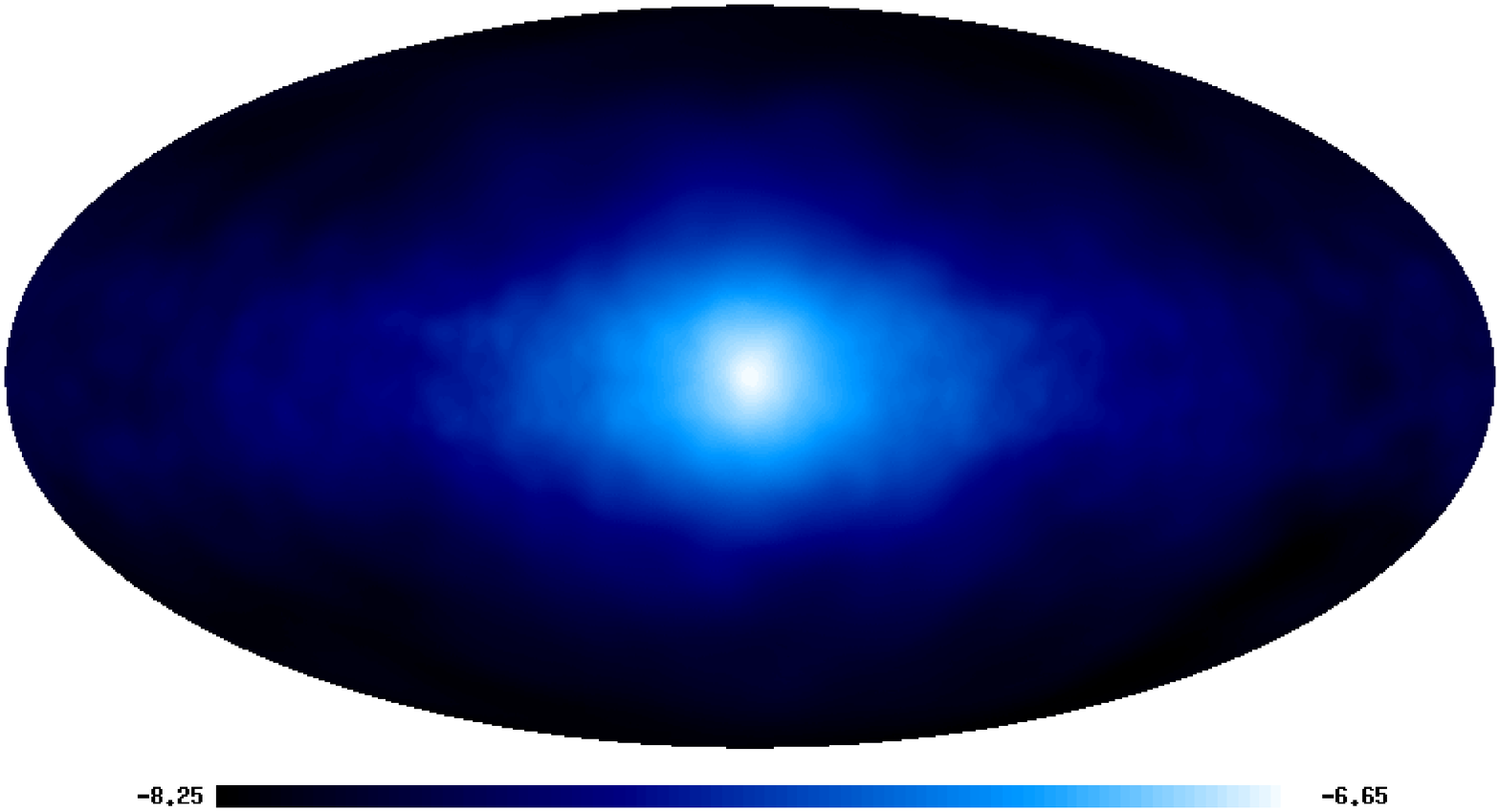, height=4.2cm}
\epsfig{file=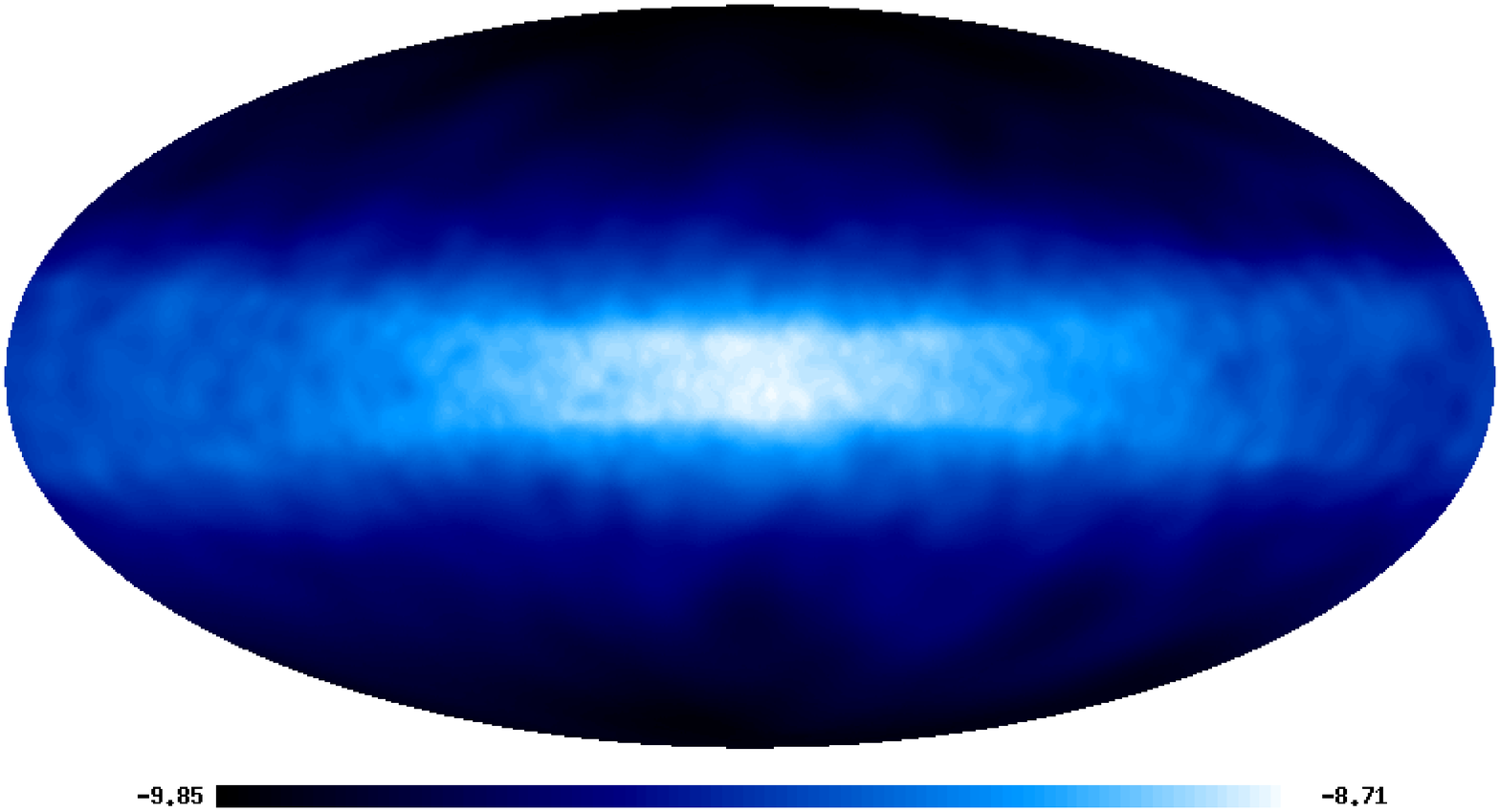,height=4.2cm}
\epsfig{file=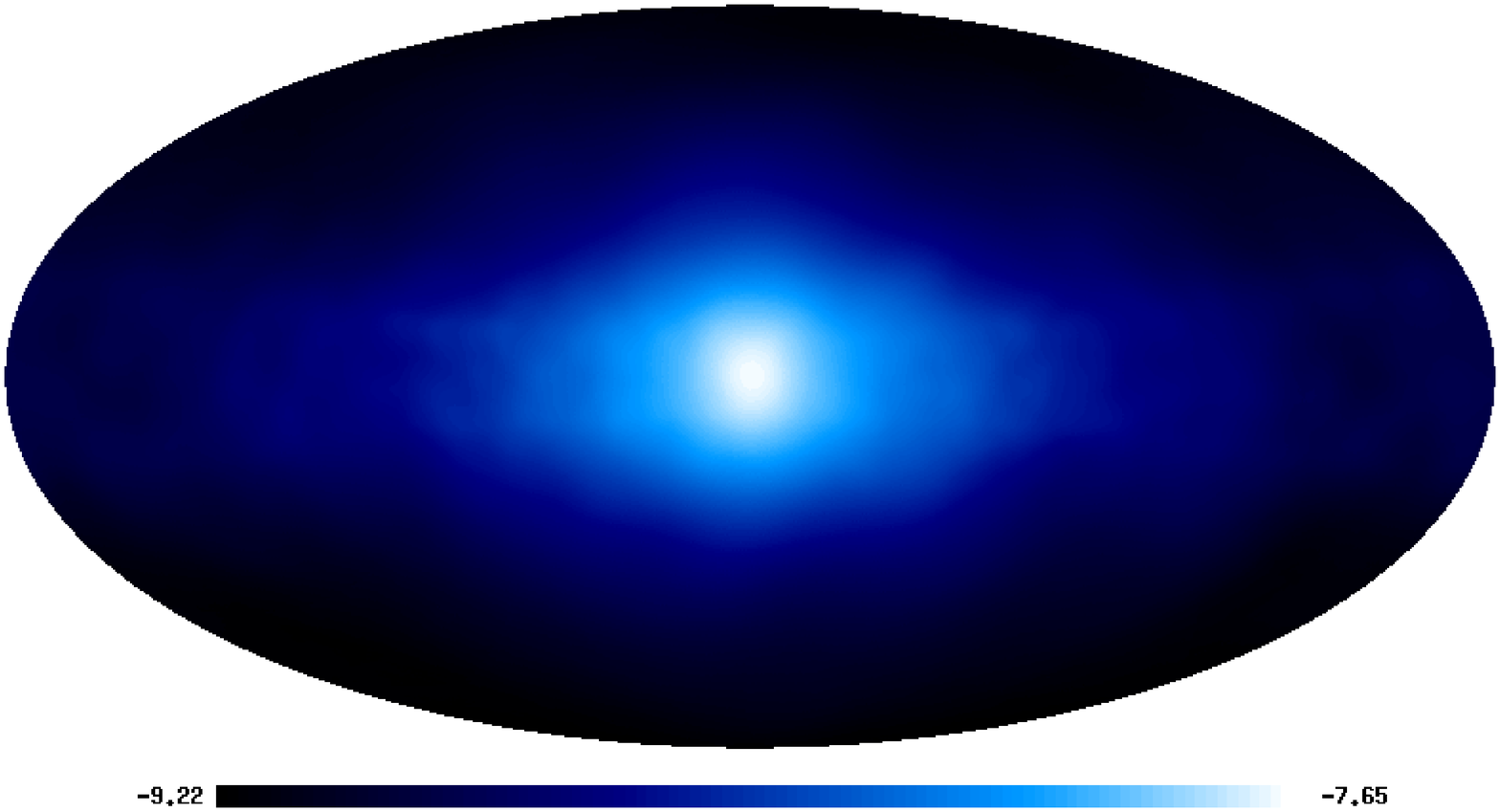, height=4.2cm}
\epsfig{file=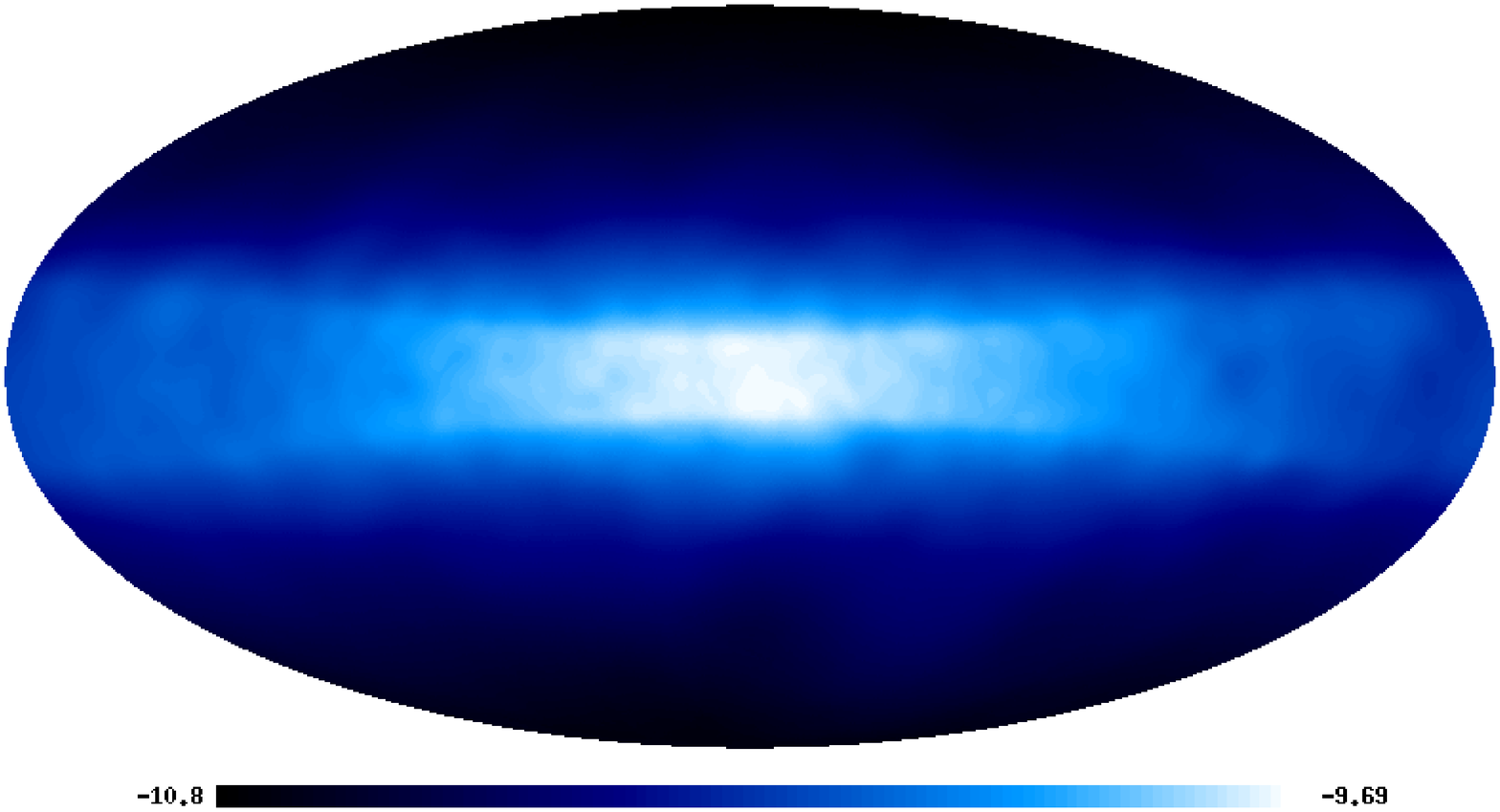,height=4.2cm}
\end{center}
\caption{\label{fig:diff-bias} Sky maps of gamma-ray emission at 1 GeV, 10 GeV and 100 GeV (from top to bottom) due to ICS from one realization of subhalos for the unbiased (left panel) and anti-biased radial (right panel) distribution. The minimum subhalo mass is $10^{-6}M_\odot$. The color scaling is logarithmic, and the unit is  $\rm 1/s/cm^2/sr$.}
\end{figure}

\begin{table}[tb]
 \begin{center}
   \begin{tabular}{|c||c|c|c|}
    \hline
     Halo model  & $E_\gamma =$ 1 GeV & $E_\gamma=$ 10 GeV   & $E_\gamma=$100 GeV   \\
     \hline
     $M_{min} = 10^{-6} M_\odot + {\rm unbiased}$   & $3.27\times10^{-7}$ & $2.03\times 10^{-8}$    &  $2.21\times 10^{-9}$   \\

\hline
     $M_{min} = 10^{-6} M_\odot + $ anti-biased  &  $8.17\times 10^{-9}$ &   $5.07\times 10^{-10}$&  $5.52\times 10^{-11}$  \\

     \hline \hline

Host halo  &  $3.83\times 10^{-10}$  &   $2.44\times 10^{-11}$  &  $2.73\times 10^{-12}$  \\

\hline

   \end{tabular}
   \caption{The mean gamma-ray intensities $\langle I\rangle$ at 1
     GeV, 10 GeV and 100 GeV, averaged over the sky, from subhalos
     with minimum mass $M_{\rm min}=10^{-6} M_\odot$ for the unbiased
     and anti-biased radial distribution and from the smooth host
     halo. The unit is ${\rm cm^2/s/sr}$. }
   \label{tab:I}
 \end{center}
\end{table}

\begin{table}[tb]
 \begin{center}
   \begin{tabular}{|c||c|c|c|}
    \hline
     $E_\gamma (\rm GeV)$  & SL ($T_p$ =3800 K) & IR ($T_p$ =40.6 K)      & CMB ($T_p$ =2.73 K)   \\
     \hline
     1    & 14 GeV (2.28 kpc)&  136 GeV (1.1 kpc)  & 526 GeV (0.48 kpc)  \\
     10    & 44 GeV (1.76 kpc)  &  431 GeV (0.65 kpc)    & 1665 GeV    \\
     100    & 141 GeV (1.26 kpc)  &  1365 GeV  & 5267 GeV \\
     \hline
   \end{tabular}
   \caption{The dependence of the characteristic electron energy $E$
     on the energy $E_\gamma$ of gamma-ray emission through inverse
     Compton scattering off the various black-body components of the
     ISRF with temperatures $T_p$. For the cases $E<1\,$TeV the
       corresponding diffusion length $\lambda_D(E)$ is also shown in
       braces.}
   \label{tab:E_e}
 \end{center}

\end{table}

In Fig.~\ref{fig:diff-unbias}, we present the sky map of gamma-ray
emission predicted by an unbiased radial distribution of subhalos:
 in order to avoid saturation from the dominant population
of small halos, in the figure we only include halos with mass above
$10^{6} M_\odot$. The most remarkable feature in
Fig.~\ref{fig:diff-unbias} is that the diffuse emission regions can
extend to a few kpc in length scale, corresponding to a few tens of
degrees on the sky. Furthermore, the size of the brightest regions
representing the largest intensity tends to increase with increasing
$E_\gamma$, which could give rise to an increase of the angular power
spectrum on corresponding angular scales. The size of the
emitting region is basically determined by the diffusion length
$\lambda_D$ which is the distance that the $e^+e^-$ diffuses during
their energy loss time. This length scale can be estimated from
Eq.~(\ref{eq:d}), giving $\lambda_D(E)\propto E^{-0.15}$ for the MED
propagation model. This energy dependence of $\lambda_D$ is shown in
Tab.~\ref{tab:E_e}. The typical diffusion length is $0.48$ kpc,
  $0.65$ kpc and $1.26$ kpc for electron energies leading to emission
  at $E_\gamma=1\,$GeV, 10 GeV and 100 GeV, respectively,
  corresponding to an angular scale $\theta\simeq\lambda_D/d$ with
  $d\sim$ few kpc the typical distance to the dark matter annihilation
  source. These angular scales are roughly what one sees in
Fig.~\ref{fig:diff-unbias} and the above estimate also applies to the
smooth host halo case (see~Fig.~\ref{fig:diff-host}) and to the
anti-biased radial distribution of subhalos.

How does the mean intensity depend on $M_{\rm min}$? Empirically, we find that the mean intensity roughly doubles with each decade of decreasing mass of subhalos, similar to what has been found by Ref.~\cite{SiegalGaskins:2008ge}. In light of the subhalo mass function $N(>M)$ in Eq.~(\ref{eq:dndm}) and the concentration parameter $c(M)$ in Eq.~(\ref{eq:c}), the annihilation rates per decade of subhalo mass can be approximated by $ L_{sub}(M) dN/d\log(M)\propto c(M)^3MN(>M)\simeq M^{-0.3} $, which is fairly consistent with our detailed numerical calculation.

In order to determine the influence of the radial distribution of subhalos on the intensity, in Fig.~\ref{fig:diff-bias} the gamma-ray full-sky maps at 1 GeV, 10 GeV and 100 GeV for the un-biased distribution are compared with the anti-biased distribution. For the anti-biased radial distribution the mean intensities are roughly 50 times smaller than for the unbiased case since the mean distance of subhalos to us is much larger than in the unbiased case. One notices the important feature in Fig.~\ref{fig:diff-bias} that the emission from the spatially anti-biased distribution is much less centrally concentrated, and apparently accumulates around the Galactic plane compared with the unbiased case. This can be understood from the fact that only electrons and positrons confined within the diffusion zone of scale height $L=4$ kpc and radius $R=20$ kpc for the MED model can efficiently produce gamma-rays by ICS and the subhalos within the diffusion zone are distributed much more isotropically for the anti-biased case compared to the more central distribution of the unbiased case. Furthermore, ICS outside the diffusion zone contributes less than $10\%$ to the mean intensity~\cite{Zhang:2009pr,Perelstein:2010fq,Cline:2010ag}.

\subsection{The angular power spectrum }\label{subsec:rcl}

\begin{figure}
\begin{center}
\epsfig{file=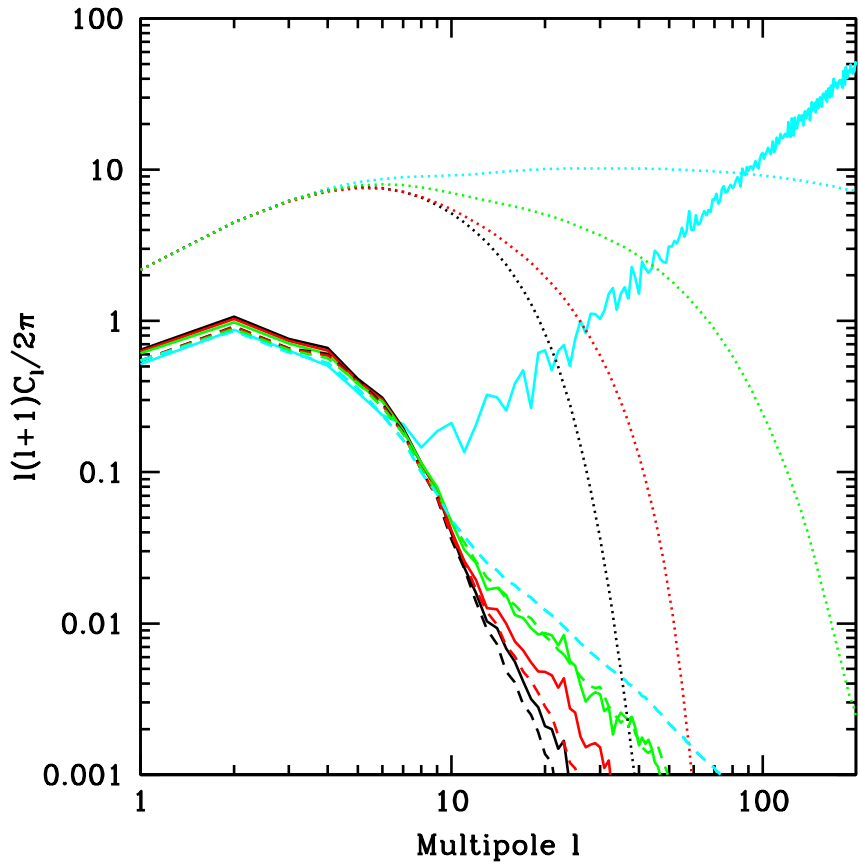, height=12cm}
\end{center}
\caption{\label{fig:cl} Dimensionless angular power spectrum $C_l$ of
  the gamma-ray sky from dark matter annihilation at $E_\gamma=1\,$
  GeV (green), 10 GeV (red) and 100 GeV (black), respectively. Solid
  curves correspond to the case of substructures with minimum subhalo
  mass $M_{\rm min}=10^{-6} M_\odot$ for the unbiased radial
  distribution. Dotted and dashed curves are for the smooth host halo
  with NFW profile, where the emissivity $\propto \rho^2$ and $\propto
  \rho$, respectively. For comparison, the cyan curves show the power
  spectrum in absence of diffusion, for the minimum subhalo
  mass $M_{\rm min}=10^{2} M_\odot$ and for the unbiased radial
  distribution (see text for details). We
  find a strong suppression due to diffusion for $l\ga10$. Each
    power spectrum is calculated exclusively from the contribution of
    the indicated source component.}
\end{figure}

\begin{figure}
\begin{center}
\epsfig{file=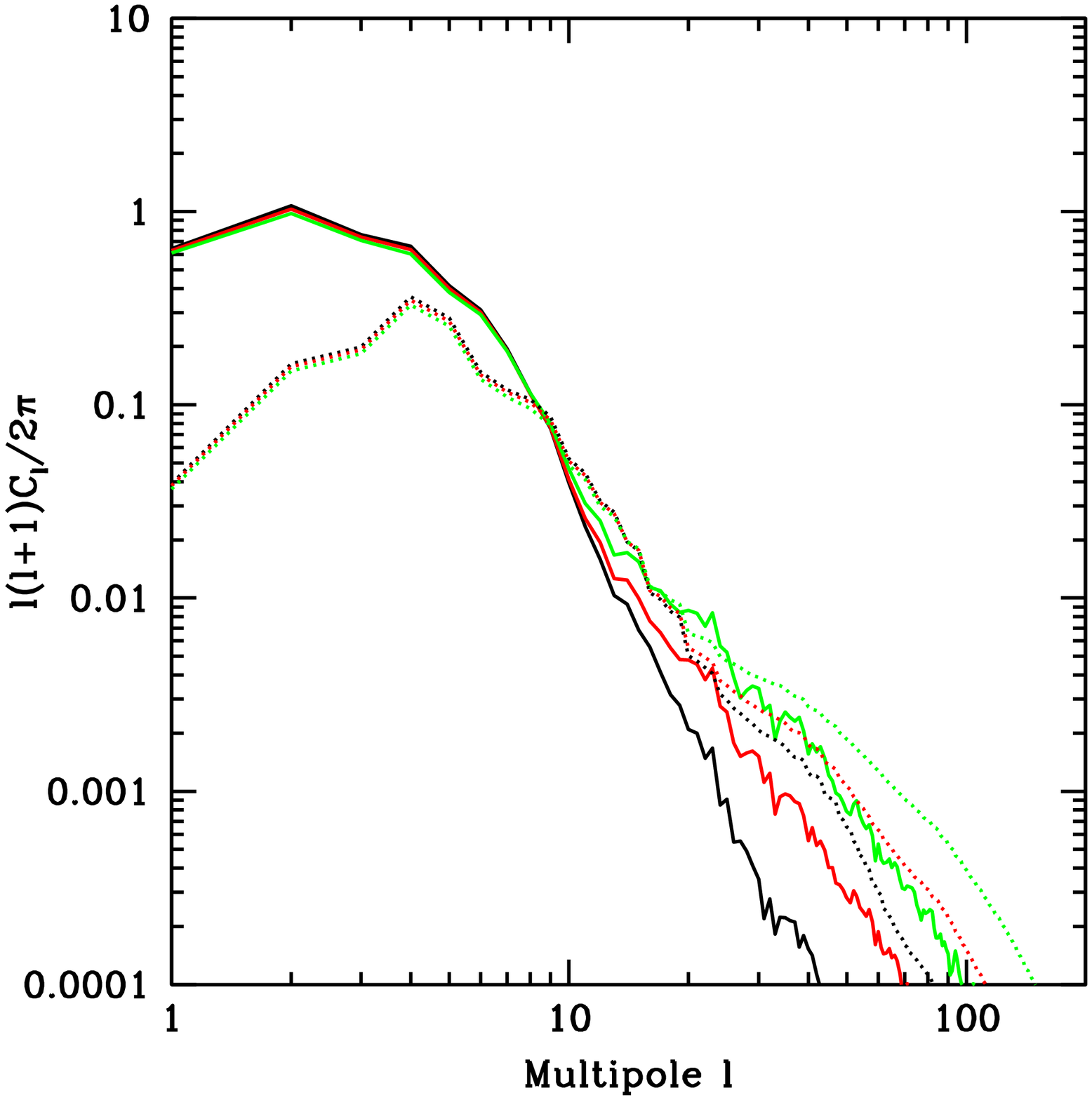, height=10cm}
\end{center}
\caption{\label{fig:bias_cl} Comparison of the dimensionless angular power spectrum $C_l$ of gamma-ray emission from dark matter substructures for the unbiased radial distribution (solid) and the anti-biased distribution (dotted) at $E_\gamma=1\,$ GeV (green), 10 GeV (red) and 100 GeV (black) with $M_{\rm min}=10^{-6}M_\odot$.}
\end{figure}

The dimensionless angular power spectra $C_l$ of gamma-ray emission
due to the ICS at photon energies 1 GeV, 10 GeV and 100 GeV from the
main components of Galactic dark matter are presented in
Fig.~\ref{fig:cl}. Shown are the power spectrum for subhalos with
$M_{\rm min} =10^{-6} M_\odot$ for the unbiased radial distribution
and for the smooth host halo. To clearly illustrate the effects of
diffusion, the angular power spectra for spatially unbiased subhalos 
with $M_{\rm min} = 10^{2}M_\odot$ and for the host halo are also
shown in absence of diffusion. We recall that in order to obtain for a
given component the contribution to the angular power spectrum $C_l^I$
in units of intensity squared, according to Eq.~(\ref{C_l}) one has to
multiply the dimensionless power spectra $C_l$ shown in
Fig.~\ref{fig:cl} by the squared total intensity $\langle I \rangle^2$
of the corresponding component from Tab.~\ref{tab:I}.

  Fig.~\ref{fig:cl} demonstrates the remarkable feature that the
  power spectrum of the ICS component of galactic dark matter
  annihilation is exponentially suppressed for $l\ga10$ compared with
  what one would obtain without diffusion. Furthermore, the lower
  energy gamma-rays have more angular power at $l\ga10$ corresponding
  to small angular scale. This can be understood from the
  energy-dependence of the diffusion length: Intensity fluctuations
  should be damped on length scales smaller than the diffusion length
  $\lambda_D$, corresponding to a multipole $l\ga\pi d/\lambda_D$,
  where $d$ is the typical distance to the dark matter annihilation
  source. Based on the discussion in Sect.\ref{subsec:dg}, we can
  estimate the diffusion length of the electrons and positrons
  emitting a given gamma-ray energy.
  When doing so, one should keep in mind that the $e^+e^-$ pairs
  interact with three different photon backgrounds and that, as it
  turns out, unlike the case of a single background, the lower the
  gamma-ray energy the higher the energy of the emitting
  electrons. Since we found that the diffusion length decreases with
the electron energy, $\lambda_D(E)\propto E^{-0.15}$, we expect 
suppression of anisotropies to occur at smaller scales, or larger 
values of the multipole $l$, for lower photon energy. For other propagation
models such as the MIN and MAX models which have a slightly different energy dependence of the spatial diffusion coefficient, the diffusion length would be slightly larger or smaller, respectively. This would shift the suppression scale in the angular power spectrum by less than a factor of two in the multipole $l$.

In fact, for the smooth host
  halo the typical distance is $d\simeq8.5$ kpc, and the suppression
  due to diffusion should occur at $l\simeq55,47$ and $28$ for
  gamma-ray energies of 1 GeV, 10 GeV and 100 GeV, respectively. 
  This analysis can be applied to the case of subhalos. For an
  unbiased spatial distribution of subhalos, the typical distance to a
  subhalo is $d\simeq7\,$kpc which corresponds to diffusive
  suppression at relatively smaller $l$ compared to the host halo
  case. These simple estimates are consistent with our detailed
  calculations shown in Fig.~\ref{fig:cl}.

  We also find that the amplitude of the dimensionless angular
  power spectrum $C_l$ from the smooth host halo is larger than that
  from the subhalos since the emissivity profile from annihilation in
  the smooth halo is proportional to the density squared and thus more
  peaked than annihilation in the subhalos which essentially follow
  the linear density profile of an NFW profile, as we see in
  Fig.~\ref{fig:cl}. We note that although the smooth host halo has a
  large amplitude of the dimensionless angular power spectrum $C_l$,
  its contribution to the total intensity is as small as $\sim0.1\%$
  for the unbiased subhalo distribution and $\sim 5\%$ for the
  anti-biased distribution, as seen in Tab.~\ref{tab:I}. We can
  therefore safely neglect the contribution from the host halo both to
  the mean intensity and to the dimensional angular power spectrum
  $C_l^I$.

Finally, we show in Fig.~\ref{fig:cl} the dimensionless angular power
spectrum $C_l$ for a smooth halo with emissivity tracing the density
of the NFW profile instead of the squared density that would be
relevant for the contribution of the host halo in decay scenarios. This shows that a smooth NFW profile describes the emission
profile by a large number of subhalos following an unbiased radial
distribution very well, at a level of $\sim0.1\%$. This is not
surprising since the number of subhalos within the diffusion zone with
masses below $10^4 M_\odot$ is sufficiently large, $> 10^{5}$, for
each mass decade to strongly suppress any deviation from a smooth
distribution. This conclusion is also true for the anti-biased
case. Despite the fact that the contribution of large subhalos in the
mass range of $10^{4}- 10^{10} M_\odot$ fluctuates strongly, their
contribution to the total emission are three orders of magnitude
smaller which leads to fluctuations at $0.1\%$ level in angular power
spectrum as seen in Fig.~\ref{fig:cl}.

How does the radial subhalo distribution affect the angular power
spectrum? In Fig.~\ref{fig:bias_cl} we compare results for the
unbiased case and the anti-biased case. At small $l$ the angular power
spectrum $C_l$ for the anti-biased distribution is suppressed relative
to the unbiased case which is due to the more isotropic subhalo
distribution at large angular scales seen in
Fig.~\ref{fig:diff-bias}. For $l\ga10$, the angular power spectrum
induced by an anti-biased distribution has more power than the
unbiased case because the typical distance to subhalos is larger for
the anti-biased distribution, shifting power to larger $l$.

We should note that, of course, the electrons and positrons of
  ordinary astrophysical sources, such as pulsars and supernova
  remnants, could also produce significant ICS signals. At
  $l\ga10$ the corresponding angular power spectrum would be
  difficult to distinguish from the one induced by dark matter because
  the large-$l$ power spectrum is highly suppressed and
  the suppression scale is mostly determined by the
  diffusion length of electrons and positrons which is independent of wether
  their origin is astrophysical or from dark matter. In contrast, the
  power spectrum at $l\la10$  should contain information on the
  sources of electrons and positrons due to the different spatial
  morphologies of dark matter and astrophysical sources on these
  scales which are not significantly influenced by diffusion. In
  addition, the ICS energy spectrum is in general significantly
  different due to the harder pair spectrum from dark matter
  annihilation   which would naturally explain the $e^+e^-$ excesses observed by PAMELA and Fermi-LAT. Although the nearby astrophysical sources such as pulsars can also reproduce these excesses, pulsars are mainly concentrated at the Galactic plane and the primary $e^+e^-$ injected spectrum have an exponential cutoff at high energies, both of which are quite different from annihilating dark matter scenarios. Therefore, a combination of these two effects may allow to discriminate between astrophysical sources and at least some dark matter scenarios. For example, Ref.~\cite{SiegalGaskins:2010mp} proposed an energy dependence of an anisotropy signature to distinguish Millisecond pulsars from heavy dark matter candidates. The disentanglement of dark matter signals from astrophysical backgrounds is, however, beyond the scope of the present paper.

\section{Conclusions}\label{sec:conclusion}
In this paper we have investigated the angular power spectrum of the
gamma-ray emission from inverse Compton scattering off low energy
target photons of electrons and positrons produced by dark matter
annihilation. We considered two extreme cases for the radial
distribution of subhalos and simulated the full-sky gamma-ray maps at
three energies through realizations of a large number of subhalos with
masses down to $10^{-6} M_\odot$. The contributions to the angular
power spectrum and to the total flux from the smooth host halo were
also calculated. 
%We then compared the predicted anisotropy signals with the Fermi-LAT observations and discussed the detectability of these signals. Our main findings are as follows:

1. We point out a new feature in the angular power spectrum of
  photons produced by inverse Compton scattering of pairs from
  annihilating dark matter that does not occur in the power spectra of
  gamma-rays produced directly in annihilation, on which earlier work
  often has focused: In contrast to the direct annihilation component for
  which the Poisson noise dominates at large $l$, the angular power
  spectrum from inverse Compton scattering is exponentially suppressed
  at $l\ga10$, because the diffusion of the high energy parent
  electrons and positrons produced by dark matter annihilation
  strongly smears out anisotropies of gamma-rays at small angular
  scales. This suppression scale is determined by the diffusion length $\lambda_D (E)$, evaluated at the energy $E$ of the emitting
    electrons and positrons. The energy $E$ of the dominant emitters is determined
    by the low energy target photon backgrounds and, in the presence of
    several background components, is not a monotonic function of the gamma-ray
    energy.  For the backgrounds used in this paper, CMB, IR and SL,
    we find that the lower energy gamma-rays are produced by the
    higher energy $e^+e^-$ pairs (and vice-versa) so that when the energy
    dependence of the spatial diffusion coefficient is taken into
    account, $\lambda_D(E)\propto E^{-0.15}$, power spectrum
    suppression is found to occur at larger multipole $l$ for lower
    photon energy.

%   In addition, decreasing gamma-ray energy corresponds to
%   increasing pair energy and decreasing diffusion length, leading to
%   in increasing multipole $l$ at which power spectrum suppression
%   occurs.

  2. The contribution to the absolute amplitude of the angular power
  spectrum from the smooth host halo can be safely neglected which is
  due to its negligible mean intensity compared to that from the
  subhalos. For subhalos, the intensity from an unbiased subhalo
  distribution is $\sim40$ times larger than from an antibiased
  distribution. At small $l$ the dimensionless angular power spectrum
  for the anti-biased distribution is suppressed relative to the
  unbiased case, whereas at large $l\ga10$ the opposite is true.

  3. An enormous number of subhalos with the masses down to $10^{-6}
  M_\odot$ can be well described by a smooth halo whose emissivity
  follows the spatial distribution of subhalos. That is not surprising,
  because diffusion of the emitting particles can smooth the whole sky map if the
  number of subhalos is sufficiently large.
  
 % 4. The optimal window to detect in the diffuse gamma-ray emission observed by 
 % Fermi-LAT the signatures of inverse Compton
 % scattering photons from dark matter annihilation into pairs
 % is at gamma-ray energies of 100-300 GeV. For $l\la10$ and in an
 % unbiased subhalo distribution, the power
 % spectrum predicted by TeV scale dark matter with a standard thermal
 % relic annihilation cross section is comparable with the observed angular power spectrum.

Finally we remark that we have assumed homogeneous and isotropic
magnetic fields and interstellar radiation field.  However, for the
more realistic case where the radiation and magnetic fields depend on
the position in the Galaxy~\cite{heiles96}, the angular power spectrum
should be modified at least at large angular scales. In particular,
the small scale fluctuations of the magnetic field could affect the
distribution of electrons and positrons~\cite{Chen:2004xx},
influencing the angular power spectrum at all scales.
Moreover, a space-dependence of the diffusion length could also 
lead to a directional dependence of the angular power spectrum.
Obviously detecting the differential effect of the anisotropy signal 
across the Galactic latitude would be even more difficult than detecting the
signal integrated over the whole sky. However, it could introduce subtle
effects when part of the sky is masked out, e.g. to avoid the contribution
from the galactic plane. These
complications should be investigated more thoroughly in the future by
performing a real 3D simulation through numerically solving the
transport equation for electrons and positrons.  While the formalism
developed in this paper can be extended to the anisotropies of the
radio sky, the small scale structure of the magnetic field is more
important for synchrotron emission and should be treated in more
detail. In addition, our work could also be extended to inverse
Compton gamma-ray emission from extragalactic dark matter halos.

\section*{Acknowledgements}
We acknowledge use of the HEALPix~\cite{Gorski2005} software and
analysis package for deriving the results in this paper. This work was
supported by the Deutsche Forschungsgemeinschaft through the
collaborative research centre SFB 676 Particles, Strings and the Early
Universe: The Structure of Matter and Space-Time, and by the State of
Hamburg, through the Collaborative Research program Connecting
Particles with the Cosmos within the framework of the
Landesexzellenzinitiative (LEXI). We thank Luca Maccione and Enrico Borriello for reading the manuscript and for very useful comments.

\end{document}